\documentclass[12pt]{aastex62}

\usepackage{multirow}
\usepackage{rotating}
\usepackage{graphicx}
\usepackage{epstopdf}
\usepackage{placeins}
\usepackage{amsmath}
\usepackage{color}
\newcommand{\RNum}[1]{\uppercase\expandafter{\romannumeral #1\relax}}

\graphicspath{{./}{figures/}}

\shorttitle{CNNs for searching superflares of {TESS}}
\shortauthors{Tu et al.}

\begin{document}

	\title{Convolutional Neural Networks for Searching Superflares from Pixel-level Data of the \\Transiting Exoplanet Survey Satellite}
	
	
	\correspondingauthor{F. Y. Wang}
	\email{fayinwang@nju.edu.cn}
	
	\author[0000-0001-6606-4347]{Zuo-Lin Tu}
	\email{tuzuolin@smail.nju.edu.cn}
	\affil{School of Astronomy and Space Science, Nanjing University, Nanjing 210093, China}
	
	\author[0000-0001-6021-5933]{Qin Wu}
    \affil{School of Astronomy and Space Science, Nanjing University, Nanjing 210093, China}
    
    \author[0000-0002-9185-4372]{Wenbo Wang}
    \affil{School of Astronomy and Space Science, Nanjing University, Nanjing 210093, China}
    
     \author[0000-0001-6545-4802]{G. Q. Zhang}
    \affil{School of Astronomy and Space Science, Nanjing University, Nanjing 210093, China}
	
	\author[0000-0002-5550-4017]{Zi-Ke Liu}
    \affil{School of Astronomy and Space Science, Nanjing University, Nanjing 210093, China}

	\author[0000-0003-4157-7714]{F. Y. Wang}
	\affil{School of Astronomy and Space Science, Nanjing University, Nanjing 210093, China}
	\affil{Key Laboratory of Modern Astronomy and Astrophysics (Nanjing University), Ministry of Education, Nanjing 210093, China}
	
	
	
	\begin{abstract}
		In this work, six convolutional neural networks (CNNs) have been trained based on 15,638 superflare candidates on solar-type stars, which are collected from the three years of Transiting Exoplanet Survey Satellite ({TESS}) observations. 
		These networks are used to replace the manually visual inspection, which was a direct way of searching for superflares, and exclude false-positive events in recent years.
		Unlike other methods, which only used stellar light curves to search for superflare signals, we try to identify superflares through {TESS} pixel-level data with lower risk of mixing false-positive events, and give more reliable identification results for statistical analysis.
		The evaluated accuracy of each network is around 95.57\%. 
		After applying ensemble learning to these networks, the stacking method promotes accuracy to 97.62\% with a 100\% classification rate, and the voting method promotes accuracy to 99.42\% with a relatively lower classification rate at 92.19\%.
		We find that superflare candidates with short duration and low peak amplitude have lower identification precision, as their superflare-features are hard to be identified. 
		The database including 71,732 solar-type stars and 15,638 superflare candidates from {TESS} with corresponding feature images and arrays, and the trained CNNs in this work are public available.

	\end{abstract}

	
	\section{Introduction}
	\label{sec:intro}
    As time-domain astronomy develops, a large amount of observational data need to be scientifically processed.
	Astronomers have applied data-mining methods including machine learning for the processing of observational data. For example, a convolutional neural network (CNN) was used to search strong-lensing systems \citep{2020MNRAS.497..556H}. Machine learning and recurrent neural networks (RNN) were used for detecting exoplanet transits \citep[e.g.][]{2021MNRAS.502.2845R,2022AJ....163...23C,2022NewA...9101693O}. \citet{2021MNRAS.503.5263Z} used random forest for the classification of four XMM data release (DR) 9 sources. \citet{2021JSWSC..11...39G} used deep-learning methods for solar flare forecasting. The unsupervised clustering method was used for classifying fast radio bursts \citep[FRBs;][]{2022MNRAS.509.1227C}. CNNs were used for searching FRBs from radio data \citep[e.g.][]{2020MNRAS.497.1661A,2021MNRAS.507.3238Y}.Machine-learning methods have already been applied to enhance the efficiency of selecting stellar flares from {Kepler} data \citep[e.g., see,][]{2018A&A...616A.163V}.

	Superflares are stellar bursting phenomena with duration and energy larger than typical solar flares \citep{Hudson2021}. The total energy of superflares on solar-type stars can be larger than $10^{33}$ erg \citep{2000ApJ...529.1026S}.
	Superflares have severe impacts on the space weather around the star and exoplanets \citep[e.g.][]{2010AsBio..10..751S, 2016NatGe...9..452A, 2017ApJ...848...41L, 2017MNRAS..465L..34A}. One solar superflare will definitely damage habitability of the Earth. The rapid increases of $^{14}$C in tree rings may be attributed to solar superflares \citep{Miyake2012,Miyake2013,Park2017,Wang2017}. Fortunately, it has been predicted that the Sun generates superflares with lower possibility \citep{2013PASJ...65...49S}.
	With the limitation of observational techniques, detailed observations of distant stars seem impossible. In recent years, astronomers try to use space telescopes to study stellar superflares. Many works have used {Kepler} data to build the connection between solar flares and stellar superflares \citep{2012Natur.485..478M, 2013ApJS..209....5S, 2015EP&S...67...59M,2019ApJ...876...58N,2021ApJ...906...72O}. 
	
	{TESS} began to observe on 2018 August 7, and covered 85\% of the sky \citep{2015JATIS...1a4003R}, while {Kepler} covered only 0.25\% of the sky \citep{2016PASP..128g5002V}.
	\citet{2020ApJ...890...46T,2021ApJS..253...35T} used {TESS} data and Large Sky Area Multi-Object Fiber Spectroscopic Telescope (LAMOST) spectroscopic data to strengthen the view that superflares can be generated by isolated stars \citep[e.g.,][]{2013PASJ...65...49S}, instead of through star-planet interactions \citep[e.g.,][]{2000ApJ...529.1031R,2004ApJ...602L..53I}.
	Comparing the statistical results from {TESS} with those from {Kepler}, it has been found that {TESS} is less sensitive for detecting relatively weaker stellar flares than {Kepler} \citep{2019MNRAS.489..437D, 2020ApJ...890...46T}. The reason is that the bandpass filter for {TESS} is 600–1000 nm \citep{2015JATIS...1a4003R}, which is longer than 420–900 nm for {Kepler} \citep{2016PASP..128g5002V}.
	
	 Superflares on other stars may give us a chance to statistically study stellar activities. The most crucial step is collecting enough true superflare events. From {Kepler} era to the {TESS} era, superflare candidates are selected through automatic algorithms and then visual inspection \citep[e.g.,][]{2012Natur.485..478M, 2020ApJ...890...46T}. Manual visual inspection requires a long time to verify true superflare events; therefore, there is an urgent demand for an automatic and efficient method to visually inspect true superflare events. One of the executable choices is machine learning. \citet{2018A&A...616A.163V} used a random sample consensus (a traditional data-mining method) to find flares in {Kepler} data. \citet{2020AJ....160..219F} used CNN to search stellar flares in {TESS}, and their network was used in subsequent research \citep{2021arXiv210907011F}.  \citet{2021A&A...652A.107V} used an RNN to search stellar flares, which can be used for both {Kepler} and {TESS} data.
	
    The above works including CNN and RNN methods enhance the efficiency and automatization of searching for stellar flares. These methods surely improve the accuracy of those collected superflares compared to other traditional light-curve outlier-searching algorithm, such as fitting the quiescent variability \citep[e.g.,][]{2011AJ....141...50W,2015ApJ...798...92W,2016ApJ...829...23D,2018ApJS..236....7H}, the iterative $\sigma$-clipping approach \citep[e.g.,][]{2019ApJS..241...29Y}, analysis of light curve changes between all pairs of consecutive points \citep[e.g.][]{2012Natur.485..478M, 2015EP&S...67...59M, 2013ApJS..209....5S, 2021ApJ...906...72O}.
	However, these works \citep{2018A&A...616A.163V,2020AJ....160..219F,2021A&A...652A.107V} just used stellar light curves to search for stellar flare events. 
	According to our experience on the visual inspection of superflare candidates from pixel-level data, there are a number of candidates showing superflare features on light curves, but without any superflare signals from pixel-level data. Obviously, this kind of candidates should not be convincingly treated as superflare events. Their superflare-like shapes in stellar light curves may be caused by CCD noises, pipeline errors or other intrusive disturbing factors. 
	The visual inspection of pixel-level data is the most direct way of excluding contamination from false-positive events. Naturally, we pursue a machine-learning method based on pixel-level data, which excludes as much false events as possible, instead of being based on stellar light curves only. Then, a database including true superflares can be constructed by machine-learning methods. Based on this database, a statistical study of stellar activities is more reliable and scientifically valuable. So, in this work, we attempt to use the deep-learning methods such as CNN and ensemble-learning methods, including stacking and voting, to search for superflare events from the pixel-level data of {TESS}. We expect that CNNs can more accurately search for true superflare events based on pixel-level data.
	
	This paper is organized as follows. In Section \ref{sec:dataset}, we briefly articulate the process of selecting solar-type stars and superflare candidates. All candidates through our improved data pipeline have been visually inspected and classified into three classes. In Section \ref{sec:CNNs}, the construction of data sets and six structured CNNs according to six feature data are given. Ensemble learning including stacking CNN and voting are also performed in Section \ref{sec:ensemble learning}. The results of sic CNNs, comparison between CNNs, and some discussions are given in Section \ref{sec:results and discussions}. A summary is presented in Section \ref{sec:Summary}.

	
\section{Solar-type stars and superflare candidates}
\label{sec:dataset}
In this section, we will briefly illustrate how to select solar-type stars and search for superflare candidates for training CNNs. For more details, one could refer to our previous works \citep[][]{2021ApJS..253...35T,2020ApJ...890...46T}. Differences and improvements will be specifically discussed in the following. 

\subsection{Selection of solar-type stars}
\label{sec:solar-type data}
We select solar-type stars according to the following criteria: (1) stellar effective temperature ($T_{\mathrm{eff}}$),  $5100 \mathrm{K} \leqslant
T_{\mathrm{eff}}<6000 \mathrm{K}$, and (2) surface gravity ($\log g$), $\log g >4.0$ \citep[e.g.][]{2013ApJS..209....5S,2015EP&S...67...59M,2020ApJ...890...46T,2021ApJS..253...35T}. These criteria are used in \cite{2012Natur.485..478M}. As the Sun is a G-type dwarf star, the effective temperature ($5100 \mathrm{K} \leqslant
T_{\mathrm{eff}}<6000 \mathrm{K}$) ensures that those selected targets are G-type stars. Surface gravity ($\log g >4.0$) ensures they are dwarf stars instead of giant or sub-giant stars. \citet{2012Natur.485..478M} also defined Sun-like stars, which are a subset of solar-type stars with more strict criteria: the effective temperature within $5600 \mathrm{K} \leqslant
T_{\mathrm{eff}}<6000 \mathrm{K}$ and stellar period longer than 10 days. Those stars whose properties are more similar to that of the Sun can also be defined as solar analogs, which belong to the subset of solar-type stars. So, the solar-type stars selected in this work are broadly similar to the Sun.
$\log g$ and $T_{\mathrm{eff}}$ are both gathered from {TESS} input catalogue \citep[TIC v8;][]{2019AJ....158..138S}.
As {TESS} has a pixel resolution of 21$''$, it may not be able to distinguish contamination apart from the main target within one pixel. The {Hipparocos}-2 catalog \citep{2007A&A...474..653V} and Gaia early data release (EDR) 3 \citet{2020arXiv201201533G} are used to mark those solar-type stars, which are possible binary systems or contaminated by brighter stars within 21$''$ distance. {Gaia} EDR3 gives $T_{\mathrm{eff}}$ estimations, with some from Gaia-DR2 \citep{2018A&A...616A...1G}. 

Specifically, from all 71,732 solar-type stars, 277 stars are possible binary systems according to {Hipparocos}-2 catalog. After crossmatching with {Gaia} EDR3 catalog, 753 stars are with brighter stars in less than 21$''$ distance, which can not be distinguished by {TESS}. 739 stars are with brighter stars in a distance greater than 21$''$ and less than 42$''$. 
Unlike what we did in our previous work \citep{2021ApJS..253...35T} where we excluded stars that are binary systems or contaminated by brighter stars, here we flag these stars in our database. 
This is because, instead of judging whether superflares are purely from solar-type stars, this work focuses on selecting superflare candidates according to pixel-level light curves with superflare shapes or features. 
This will also enhance the purpose of collecting as much data as possible in order to obtain better performance of CNNs. Besides, 5435 stars may contain M-type star candidates within 42$''$ range. \citet{2021MNRAS.502.2033J} found that the flares from M-type stars neighboring the {TESS} targets occupy a 5.8\% $\pm$ 1.0\% rate of flares in the {TESS} data. These stars are recommended to be carefully inspected in the catalog. As {Gaia} EDR3 does not include $\log g$, and Section 3.1 of \citet{2021ApJS..253...35T} demonstrated that observations of solar-type stars may not be influenced by M-type stars, these stars are not excluded but marked with flags. Totally, 71,732 solar-type stars from three years of {TESS} observations are collected. 

We use pre-search data conditioned (PDC) light curve of every single star for period estimation and photometric variability calculation. The periods of these stars are estimated through the Lomb-Scargle method \citep{1976Ap&SS..39..447L,1982ApJ...263..835S}. We also set the false-alarm probability to $10^{-4}$ to estimate the periodicity of a star \citep[e.g.,][]{2019MNRAS.489.5513C}. The photometric variability ($R_{\rm var}$) is derived by 
\begin{equation}
	\label{equ:rvar}
	R_{\rm var} = F_{95\%} - F_{5\%},
\end{equation}
where $F_{95\%}$ and $F_{5\%}$ stand for the upper 95\% and lower 5\% of the ranked normalized flux of stellar light curves \citep{2010ApJ...713L.155B}. The first- and third-year observations of {TESS} both cover the southern hemisphere of the sky. Stars in this area are observed multiple times, so their period and $R_{\rm var}$ are calculated based on all observations from {TESS}. Note that, combining all observations from {TESS} is accomplished by stitching light curves from different sectors of {TESS} data. We used a PDC light curve of every single star from each sector and normalized the flux within the sector (about 27 days), then stitched the normalized flux from different sectors. Due to the unique observational mode of {TESS}, some stars are only observed for 27 days. Even if we stitched light curves from multiple sectors, the estimation of longer stellar periods would be hard due to the limitations of {TESS} \citep[][]{2021ApJS..253...35T}. Other methods should be considered for determining stellar periodicities of {TESS} stars, for instance, measuring the rotational velocity from the stellar spectrum \citep[e.g.,][]{2008ApJ...684.1390R, 2012AJ....143...93R}, or using the autocorrelation function \citep[e.g.,][]{2014ApJS..211...24M}. Specific information of 71,732 solar-type stars can be found in Table \ref{tab:solar type stars}.

\subsection{Superflare candidates}
\label{sec:superflare candidates}
\begin{figure*}[ht]
	\centering
	\includegraphics[width=1\linewidth]{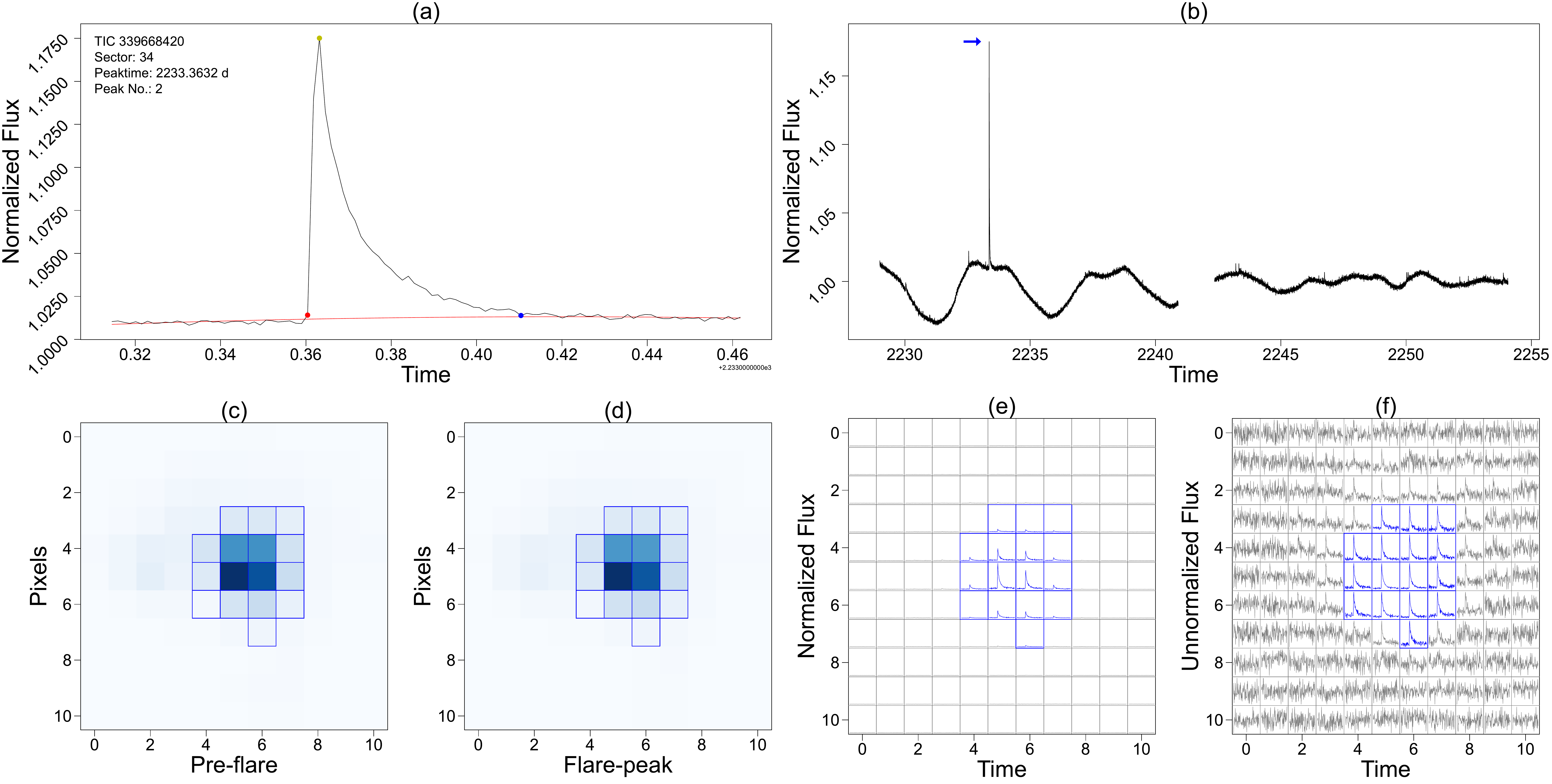}
	\caption{Patterns for visually inspecting superflares. The light curve of a superflare candidate is shown in panel (a). Here, $F(t)$ in Equation \ref{equ:fflare} is represented by the black solid line. $F_{\rm q}{(t)}$ is shown by the red solid line. Yellow, red and blue dot circles represent the peak, start, and end time stamps of this superflare, respectively. Panel (b) gives the whole light curve of the sector, from which this superflare candidate is searched. The peak of this superflare is pointed by the blue arrow mark. Panels (c) and (d) show the pixel-level images at the pre-flare and peak time stamps of the superflare, respectively. In these two panels, blue frames encircles the aperture masks of {TESS} pipeline. We use these two panels to help us check if the area of apertures are stable and unchanged, and if the brightest area is identical before and at peak of the flare. If it changes, the quality of the data will be specially considered. For each pixel in the images, its light curves with normalized and unnormalized flux are correspondingly shown in the square frames of the panel (e) and (f). Aperture masks by {TESS} pipeline are shown by blue frames in these lower panels.}
	\label{fig:TIC 339668420_34}
\end{figure*}

We select superflare candidates through the same method as in \cite{2020ApJ...890...46T,2021ApJS..253...35T}. Each solar-type star' PDC light curve from {TESS} has been checked by the automatic algorithm. The updated algorithm has been used to process {TESS} second-year observation \citep{2021ApJS..253...35T}. Briefly, those outliers in the light curve are selected by the following equation,
\begin{equation}\label{equ:deltaF}
	\Delta F^{2}\left(t_{i,
		n}\right)=s\left(F_{i}-F_{i-n-1}\right)\left(F_{i+1}-F_{i-n}\right),
\end{equation}
where $F$ is the normalized flux of one sector. $s=1$, when $\left(F_{i}-F_{i-n-1}\right)>0$ and $\left(F_{i+1}-F_{i-n}\right)>0$ are both satisfied; otherwise $s=-1$. As $n$ describes the rising time of flares \citep{2020ApJ...890...46T}, we set $n=2$ and $n=5$ to search outliers on the short and long rising phase of light curves, respectively. If $n=2$, we cut the light curve 0.01 to 0.05 days before, and 0.05 to 0.10 days after the peak flux for further analysis. Correspondingly, if $n=5$, we use the light curve with intervals of 0.03 to 0.15 days before, and 0.15 to 0.25 days after the peak flux. The quadratic function $F_{\rm q}{(t)}$ is used to fit these light curves, in order to deduct the long term variability of the star.
%
We can obtain the light curve, which keeps flare signals, through the following equation,
\begin{equation}\label{equ:fflare}
	F_{\text {flare }}(t) = F(t)-F_{\rm q}{(t)}.
\end{equation}
The first and last points, whose photomeric errors $F_{\rm error}(t)$ satisfy the following equation, are set to be the start and end time of a flare,
\begin{equation}\label{equ:errflux}
	3\times F_{\rm error}(t) < F_{\rm flare}(t).
\end{equation}
$F_{\text {flare }}(t_{\rm peak})$ at peak point stands for the peak amplitude of a flare. The stellar bolometric luminosity is calculated through the Stefan-Boltzmann law
\begin{equation}\label{equ:lumi}
	L_{*}=4 \pi R_{*}^{2} \sigma_{\mathrm{sb}} T_{\rm eff}^{4},
\end{equation}
where the stellar radius $R_{*}$ and surface temperature $T_{\rm eff}$ are taken from TIC v8 catalog. 
We estimate the superflare energy by the equation  
\begin{equation}\label{equ:flareenergy}
	E_{\text {flare }}=\int L_{*} F_{\text {flare }}(t) d t,
\end{equation}
where the stellar luminosity and flare flux are calculated in Equations (\ref{equ:lumi}) and (\ref{equ:fflare}), and the integral is from the start to the end time of a flare. Totally, 15,638 flare candidates are selected from all solar-type stars in the three years of {TESS} observations. They are all visualized through the images as shown in Figure \ref{fig:TIC 339668420_34}.
It should be noted that we standardize the time intervals, which start from 0.05 days before and 0.1 days after the flare peak time stamp. The peak time stamp is just at the 1/3 position of the light curve. This process ensures that the slowly decaying phase of superflares can be presented in the images. Panels (c) and (d) of Figure \ref{fig:TIC 339668420_34} show the star original pixel-level images before the flare and at the flare peak, respectively. The blue frames encircle the aperture masks of the {TESS} pipeline, which should be the same in the panels. Panels \ref{fig:TIC 339668420_34}(e) and \ref{fig:TIC 339668420_34}(f) show pixel-level light curves with the normalized and unnormalized flux, respectively.

\begin{figure*}[ht]
	\centering
	\includegraphics[width=1\linewidth]{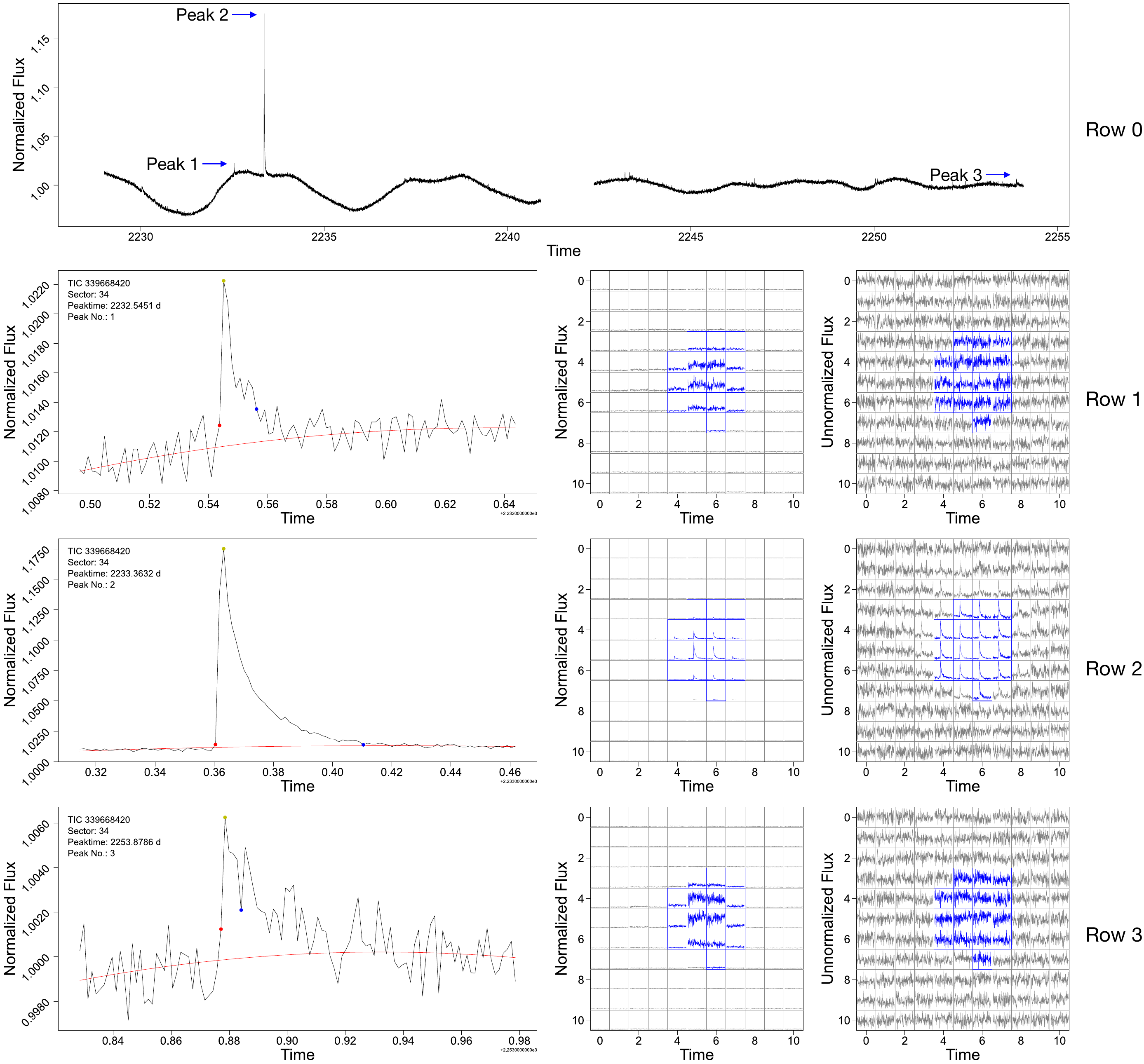}
	\caption{Examples of data in the gold class, silver class, and none class. The panel in the row 0 shows the whole light curve of TIC 339668420 in Sector 34 of {TESS}. Specific light curves of peaks 1, 2 and 3 are displayed in rows 1, 2 and 3, respectively. Corresponding pixel-level light curves are also shown besides, with both normalized and unnormalized fluxes. The red solid lines stand for $F_{\rm q}{(t)}$. The colored three dots stand for the start, peak, and end time stamps of the flare. The patterns contain pixel-level light curves, which are exactly same as those shown in Figure \ref{fig:TIC 339668420_34}. Here, peak 1 and peak 2 stand for candidates that belong to the silver class and gold class. Peak 3 belongs to the none class as a nonsuperflare (negative) event.}
	\label{fig:GoldSilverNone}
\end{figure*}

We classify all candidates into three classes, through visual inspection of their pixel-level information according to the images shown in Figure \ref{fig:TIC 339668420_34}. 
These three classes are the gold, silver class and none class. 
\begin{itemize}
\item The gold class collects those candidates which show standard superflare features with rapid rising and slow decaying shapes in the light curve. From pixel-level information, superflare signals show shapes of a standard point-spread function (PSF) as shown in the panel (e) and (f) of Figure \ref{fig:TIC 339668420_34}, where flare signals close to the stellar center pixel are obviously stronger than other pixels nearby. Almost all pixels in the {TESS} aperture show superflare signals as shown by the panel with the unnormalized flux. The patterns in row 2 of Figure \ref{fig:GoldSilverNone} are the same as those in Figure \ref{fig:TIC 339668420_34}, from which this candidate is convincingly classified as the gold class.

\item The silver class gathers superflare candidates that do not perfectly show flare features unlike those in the gold class. The silver class may also show flare signals from pixel-level data of the normalized flux. For instance, pixel [5,5] and [6,5] (where [horizon index, vertical index]) in row 1 of Figure \ref{fig:GoldSilverNone} show pinnacle-shaped light curves both in panels with normalized and unnormalized fluxes. But the pixels around them show higher noise than those in the gold class.
An example is shown in the row 1 of Figure \ref{fig:GoldSilverNone}.

\item The none class collects other candidates, which are false events including planets transit and cosmic ray signals. Row 3 of Figure \ref{fig:GoldSilverNone} gives an example of the none class. We hardly find pinnacle-shaped light curves in pixel-level data. Even if a few light curves with unnormalized fluxes show a pinch of flare-shape features, unrelated noise are strongly dominant.
\end{itemize}
To be clear, the candidates in the gold class and silver class are true superflares. The reason we split these superflares into two categories is that they have different features. In particular, for data in the silver class, bright pixels (e.g., pixel [5,5] and [6,5] ) in the panels with pixel-level light curves in Figure \ref{fig:GoldSilverNone} show pinnacle-shaped light curves, (see comparison in row 1 and row 2 of this figure). These pixels should be more cautiously inspected, because they show superflare signals but are not perfect as those in the gold class.

All 15,638 candidates are classified into these three categories by manual visual inspection. Detailed information can be found in Table \ref{tab:candidates}. For the convenience of users who are only interested in data of superflares, the duration, peak luminosity and energy of superflares are collected in this table. This information will be concealed for those candidates if at least one of following criteria is satisfied. (1) The hosting stars are with flags HB or GB21 in Table \ref{tab:solar type stars} after cross matching with the {Hipparocos}-2 and {Gaia} EDR3 catalogs. (2) The candidates' rising epoch is longer than the decaying epoch. (3) Candidates are visually inspected and classified in the none class. A column in Table \ref{tab:candidates} lists all the file names that correspond to feature images and arrays used for CNN training and will be articulated in Section \ref{sec:features_pics}.

\section{Deep learning}
\label{sec:CNNs}
This section will articulate the process of obtaining three-dimensional arrays and feature pictures for training CNNs, and how we set training, validation and test data sets. Specific structures of six networks are also shown. Two methods of ensemble learning are used for enhance performances of CNNs, including stacking and voting.

\subsection{Three-dimensional arrays and feature images}\label{sec:features_pics}
This work focuses on automatically identifying superflares through the pixel-level data of {TESS}. Figure \ref{fig:TIC 339668420_34} and Figure \ref{fig:GoldSilverNone} reveal pixel-level light curves, according to which we visually classified those superflare candidates into three classes.

\begin{figure*}[ht]
	\centering
	\includegraphics[width=1\linewidth]{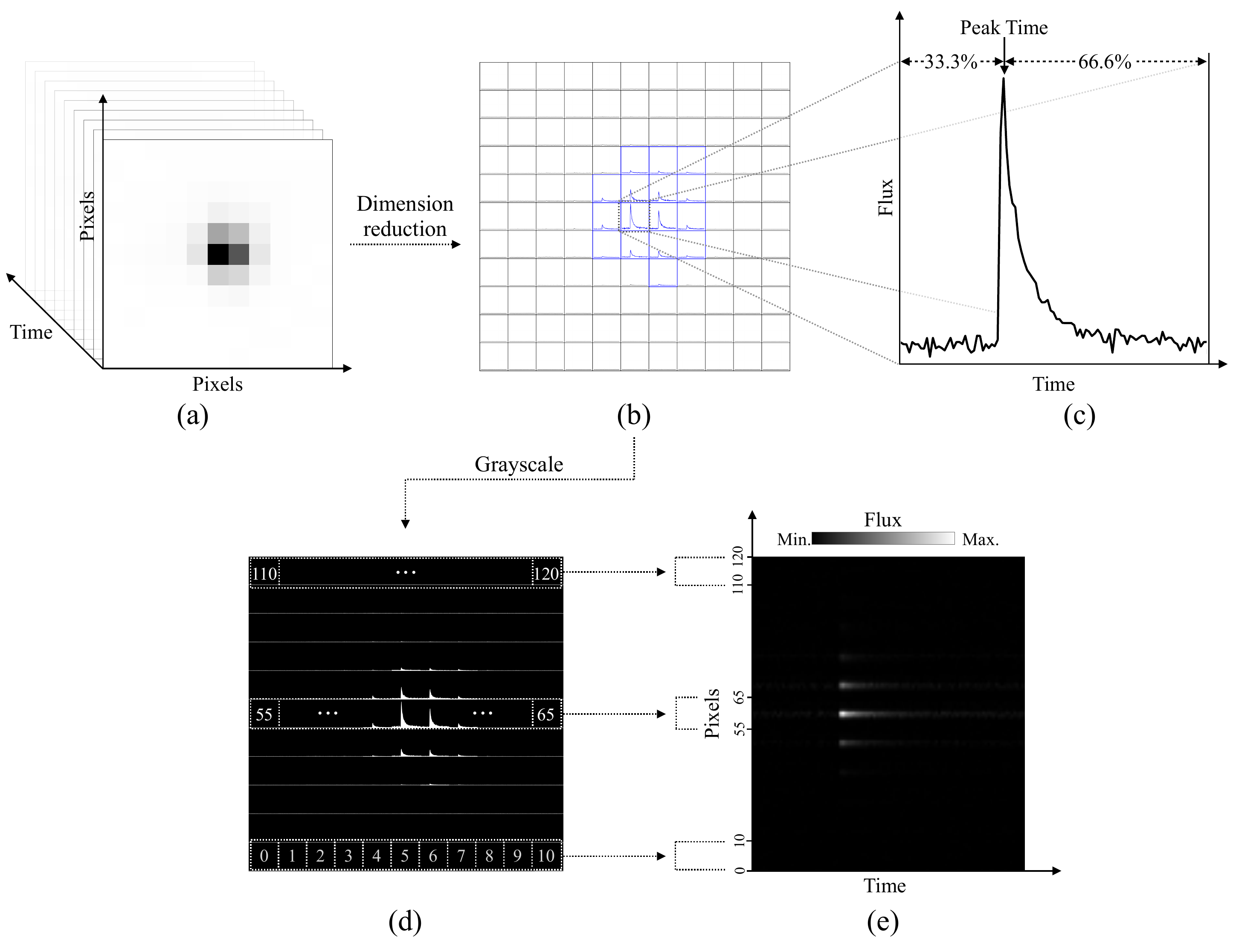}
	\caption{(a) Visualization of three-dimension array of pixel-level data. The data cover the whole flare time interval from 0.05 days before to 0.10 days after the flare peak. All time stamps from original {TESS} data are kept. Each time slice represents the pixel-level image at a specific time stamp.
	(b) Dimension reduction of the data in panel (a). We derive the light curve of each pixel in the image and reveal the light curve in the square. The square corresponds to the exact pixel site in the pixel-level image. This panel is exactly the same as panel (e) and (f) of Figure \ref{fig:TIC 339668420_34} with normalized flux. 
	(c) Zoom in one pixel light curve in panel (b). The $x$-axis and $y$-axis represent the time and flux, respectively. The peak of a candidate is set at the 33.3\% ($1/3$) position of the whole light curve. 
	(d) Convert pixel-level light curves to the image with gray scale color. Blue frames in panel (b) represent the masked apertures by {TESS}, in which light curves are correspondingly colored in white in panel (d). Other light curves are colored in grey. The areas under the light curves are filled, which is for protruding the pinnacle shape of the superflare light curve. This is an example image of PL-normalized. 
	(e) Example image of LS-normalized. We label all pixel light curves from left to right and from bottom to top as channel 0 to 120 in panel (d). Light curves displayed in each channel can be treated as an one-dimensional array, of which each value represents the flux and the index sequences represent the time series. In panel (e), we transform the way of displaying these data by lining up all sequences of channels from 0 to 120 as $y$-axis. The $x$-axis represents time series. The flux is presented by brightness of gray scale color. So, each horizontal line of this image stands for an array data displayed as a pixel light curve in the corresponding frame of the image shown in panel (d). 
	Pixel light curves with unnormalize flux (panel (f) of Figure \ref{fig:TIC 339668420_34}) are also transformed through the above procedures. Examples of PL-unnormalized and LS-unnormalized can be found in Figure \ref{fig:exa_PL_LS}. }
	\label{fig:feature_pics}
\end{figure*}

In particular, we transform pixel-level data into arrays and feature images for further CNN training. The first data format is a three-dimensional array as visualized in Figure \ref{fig:feature_pics}(a). One dimension of this data is time, and the other two are pixel indexes. This array contains original pixel-level information during the whole flare light curve interval. One may comprehend these arrays as short film clips that record changes in the pixel-level image in just 0.15 days. The image from the {TESS} data contains 11$\times$11 pixels and with 2 minutes cadence. So, these three dimensional arrays are of size 11$\times$11$\times$108. These arrays contain original observations of flare candidates from {TESS}.

Then, we transform the array into pixel-level light curves as shown in Figure \ref{fig:feature_pics}(b), which is exactly the same as that shown in panels (e) and (f) of Figure \ref{fig:TIC 339668420_34}. In this feature picture, we list 11$\times$11 squares, which correspond to 11$\times$11 pixels of {TESS} pixel-image data. By this procedure, we reduce the three-dimensional data into two-dimension images. In each square, the light curve of the corresponding pixel is revealed in Figure \ref{fig:feature_pics}(c). In order to reduce complexities of the following CNNs, we set the peak time of each pixel-level light curve at the 1/3 position of the 
time axis. Besides, the blue frames stand for those aperture masked by {TESS} pipeline. In order to reduce the color information, which is totally unrelated with superflare searching, the feature image is transformed to a gray scale image with just one channel (traditional colorful images are with RGB three channels), as shown in Figure \ref{fig:feature_pics}(d). In order to protrude the pinnacle shape of superflare light curves, the areas under the light curves are filled. We use white color for those light curves, which are from those aperture pixels masked by {TESS}, so they are distinguished from other ordinary light curves, which are in gray.

Further, we line up the sequences of all 11$\times$11 pixel light curves and reveal the flux in gray scale color as shown in Figure \ref{fig:feature_pics}(e). Each line of this feature image stands for the corresponding pixel, and from left side to right side it stands for the time axis of the flare light curve. For example, in the image, the line of pixel No. 60 suddenly brightens and gradually darkens, which means the rapid rise and slow decay of the light curve.

\begin{figure*}[ht]
	\centering
	\includegraphics[width=1\linewidth]{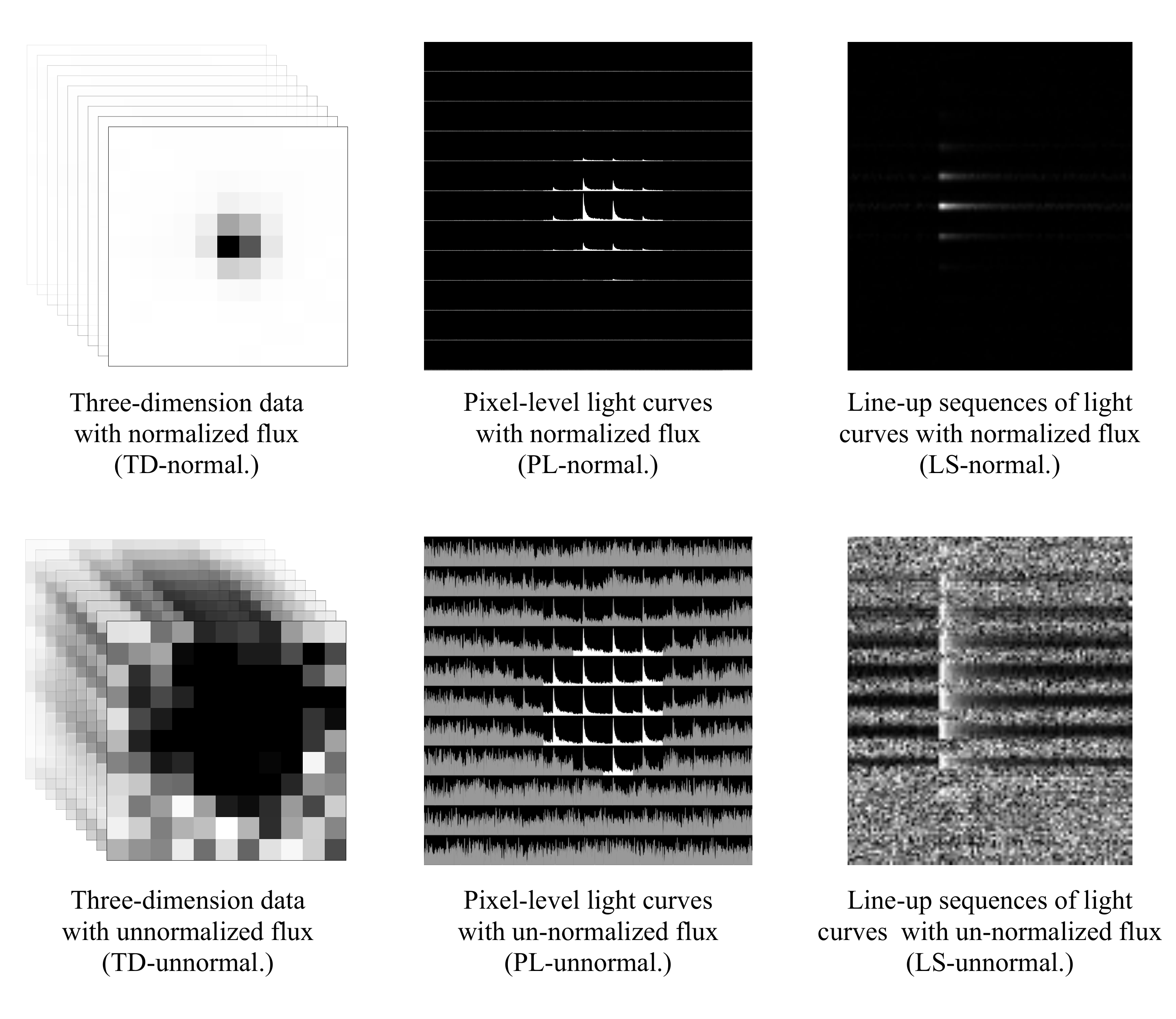}
	\caption{Examples of TD arrays, PL images, and LS images, which are all derived from the candidate displayed in Figure \ref{fig:TIC 339668420_34}. Here, normal and unnormal represent whether flux data are normalized and unnormalized. 
	Figure \ref{fig:feature_pics} illustrates how we get these feature images. 
	Examples of TD-normal. and TD-unnormal. may not be accurately visualized. 
	However, these two kinds of data are three-dimensional arrays with one dimension describing time. 
	These arrays can be understood as film clips, which reveal changes in the pixel-level image during the whole flare times. In PL-normal. and PL-unnormal. images, flare light curves which are colored in white are those data from apertures masked by {TESS}. Other pixel light curves are colored in gray.}
	\label{fig:exa_PL_LS}
\end{figure*}

As the unnormalized and normalized fluxes are both derived, we have two versions for three-dimensional arrays; images of pixel-level and lineup sequences of light curves. These six kinds of data are visualized in Figure \ref{fig:exa_PL_LS}. Then we construct six CNNs for all of them. 
All of these six feature images and arrays are from a superflare candidate. In other words, we intend to use the six CNNs to classify just one superflare candidate, because these feature images and arrays describe a candidate from different angles even according to the same pixel-level information from {TESS}.
\begin{itemize}
\item Three-dimensional arrays (TD arrays) are basically original records of {TESS} observations but without any details on the {TESS} aperture masks. 
\item Pixel-level images (PL images) are with information of the {TESS} aperture masks and show the pixel-level PSF shape of a true superflare, which shows the relationships between stellar center pixel and nearby pixels. 
\item The images of the lineup sequence light curves (LS images) more obviously reveal the contrast of dark and bright, which more delicately portray the rapid rising and slow decaying phases of superflare light curves. However, the LS images deficient in any information of {TESS} aperture masks.
\end{itemize}

\begin{figure*}[ht]
	\centering
	\includegraphics[width=1.\linewidth,angle=-90]{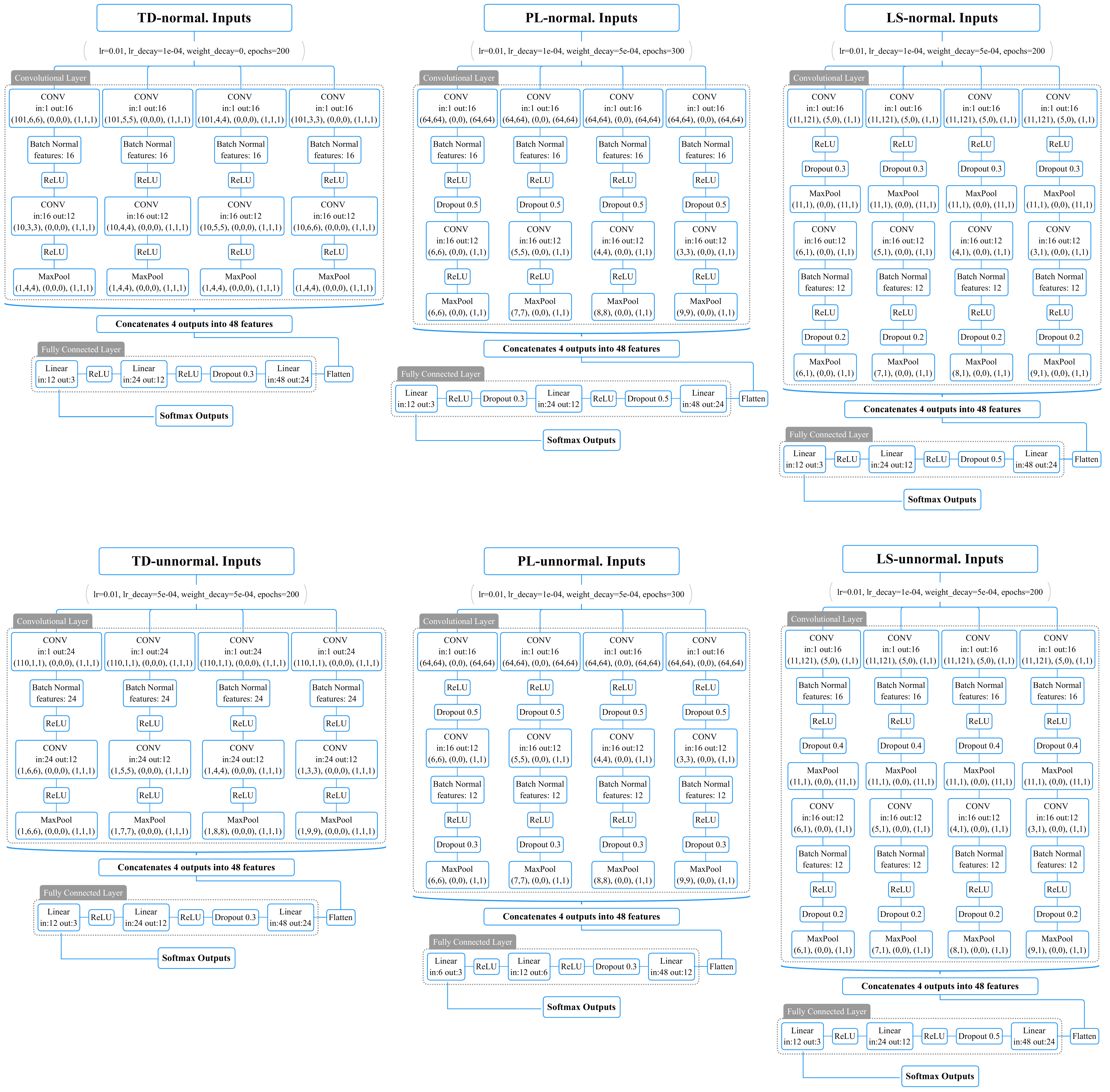}

	\caption{Flow charts of six CNNs. For each network, its learning rate (lr), learning rate decay (lr\_decay), weight decay, and epochs are all listed under the title of the network. 
	All networks are constructed by convolutional layer and fully connected layer. The three brackets in some layers denote the neural kernel size, kernel moving stride, and padding size. 
	The numbers of the input channels and output channels are also listed in the layer. The outputs of each network are applied to the Softmax function (Equation \ref{equ:softmax}), so outputs can be regarded as probabilities for classifying a candidate into three classes. }
	\label{fig:CNN-flow}
\end{figure*}

Data augmentation is a method that can be used to augment the size of the data set and make the data be rich in variation. We do not apply this method to the feature images for the following reasons.
\begin{itemize}
\item The pixel-level light curves, as shown in panels (e) of and (f) in Figure \ref{fig:TIC 339668420_34}, should reveal rapid rising ahead of slow decaying as shown by the typical superflare shape. This unique shape is what we obtain from superflare signals and also a vital factor when we inspect superflare candidates. So, any augmentation methods like flip or rotate will totally change this basic factor. Further, we may get a wrong CNN to recognize false superflare signals with slowly rising inversely ahead of rapid decaying. This situation should be avoided.
\item Superflare signals from light curves can be treated as outliers. Selecting those outliers is successfully implemented, as in Section \ref{sec:superflare candidates}. Similar methods can also be executed efficiently according to other outlier-selection algorithms \citep[e.g.,][]{2011AJ....141...50W, 2015ApJ...798...92W,2019ApJS..241...29Y}. In this work, CNNs are used to recognize feature images exported by the algorithm. As this algorithm can automatically select outliers on light curves and process pixel-level data of TESS with outputting unified feature images and arrays, CNN structures can be as lightweight as possible (Figure \ref{fig:CNN-flow}). Lightweight CNNs also ensure that we can train them more easily with limited data.
So, data from any augmentation methods like image zooming, padding seems overwhelming for the lightweight CNNs in this work.
\end{itemize}

\subsection{Training, validation, and test data sets}\label{sec:cnn datasets}
As we introduced in Section \ref{sec:superflare candidates}, 15,638 flare candidates are selected and classified into three classes including 1268 data in the gold class, 3792 data in the silver class, and 10,578 data in the none class. For convenience of the further $k$-fold cross validation of CNNs, we randomly split 15,638 flare candidates into seven stacks. Figure \ref{fig:CNNdataset} shows the workflow of this process. Each of the six stacks contains 2,348 data, and the last stack contains 1550 data. Basically, the test set takes 15\% of the whole data set. From these stacks, we choose each of them as the test data set, and others are integrated and then evenly split into 10 ministacks. One of the ministacks is set as the validation data set; the others are combined as the training data set. The validation set is about 10\% of the whole database, and the training set is 75\%. According to above flows, one test data set corresponds to 10 validation and 10 training data sets. As there are no intersection data among the test, validation and training data sets, we obtain 70 clusters of data (7 test sets $\times$ 10 training and validation sets). Each one includes a training set, a validation set, and a test set. As these three classes are data imbalanced, we try to ensure the above stacks and ministacks contain data of three classes with almost the same proportions to avoid issues due to the different proportions in $k$-folding cross validation.

\subsection{Convolutional neural networks}\label{sec:six cnns}
In this work, selecting superflares from light curves can be considered as a time-series problem or time-series forecasting problem, which is traditionally solved using RNNs instead of CNNs \citep[e.g.,][]{HEWAMALAGE2021388, 2021RSPTA.37900209L}. There are two reasons for choosing CNNs in this work. First, RNNs can be used to process light curves of stars \citep[e.g.,][]{2018A&A...616A.163V,2021MNRAS.502.2845R,2022AJ....163...23C,2022NewA...9101693O}, but selecting superflare signals can be treated as finding outliers, which can be easily executed by enormous simple but strong algorithms as we introduced in the Section \ref{sec:intro}. So, it is unnecessary for using RNNs to process stellar light curves. 
Second, according to our experience of manual visual inspection for superflare events, through only one univariate stellar light curve may lead to false-positive results, which can be avoided by using the additional information provided by pixel-level data. For instance, the left panel of row 3 in Figure \ref{fig:GoldSilverNone} shows a rapid rise and slow decay in the univariate light curve. However this event is a nonsuperflare event, as there are no superflare signals from pixel-level light curves shown in the middle and right panels of row 3 in Figure \ref{fig:GoldSilverNone}. As shown in panels (e) and (f) in Figure \ref{fig:TIC 339668420_34} and row 2 of Figure \ref{fig:GoldSilverNone}, from the PSF shape it can be found that the brightest pixel shows a higher superflare amplitude than other surrounding pixels, and outer pixels show stillness over time. Besides, the aperture masks of {TESS} pipeline are shown as blue frames in these figures, which also provide pixel-level information, and ensure that superflare signals are revealed inside of these masks, as we introduced in Section \ref{sec:features_pics}. This information can be easily identified through CNNs rather than RNNs as the information is also variable in the pixel-level data as {TESS} target changes. So, CNNs are practicable because we transform a time-series problem into an image-classification problem. Those images contain much more pixel-level information, which is helpful and necessary for identifying true superflare events. 

We have set the training, validation and test data sets; then we construct six CNNs according to the candidates' TD arrays, PL images and LS images. The validation data set is used to modulate the hyperparameters of the networks. Before we select the six CNNs for each of the TD- PL- LS-normal. and unnormal. data, we should construct a whole bunch of networks and select the most stable and outstanding one according to the validation data set. 

Flow charts of these six CNNs are shown in Figure \ref{fig:CNN-flow}. We use AdaGrad algorithm \citep{duchi2011adaptive} as an optimizer when training the networks. The learning rate, learning rate decay, and weight decay are all listed in the figure for each of these networks. None of these networks are the same, as we consider different identities of feature images and arrays. The stability shown by the validation data after applying for each network is also regarded. Note that, in order to standardize the inputs of each network, we resize PL images and LS images with resolution of 704$\times$704 and 121$\times$121, respectively. The TD arrays are all resized to 11$\times$11$\times$110 through interpolation along time-axis. We use the outputs, to which we apply the Softmax function (normalized exponential function \citet{2006Bishop}) as follows
\begin{equation}\label{equ:softmax}
	c_{i}=\frac{e^{o_{i}}}{\sum_{j=0}^{K} e^{o_{j}}} \quad \text { for } i=0, \ldots, K.
\end{equation}
As we only have three classes, $K=2$. $o_{i}$ represents the original output for the ${i}^{\rm th}$ class. So, after being exponentially normalized, $c_{i}$ represents the possibility of classifying one input candidate to ${i}^{\rm th}$ class. We choose the maximum number of $c_{i}$ to classify the input into ${i}^{\rm th}$ class. During the training of the networks, the cross-entropy loss function is used, which can be derived as 
\begin{equation}\label{equ:crossentropy}
	\ell=-\frac{1}{n} \sum_{j=1}^{n} \sum_{i=0}^{K} t_{i}^{(j)} \log {o}_{i}^{(j)},
\end{equation}
where $K=2$, and $n$ stands for the number of class and counts of data in the training set, respectively. $j$ means the ${j}^{\rm th}$ data of the training set. $t_{i}=1$ if the ${j}^{\rm th}$ data are manually classified into the ${i}^{\rm th}$ class; otherwise $t_{i}=0$. The purpose of training the networks is to reduce the value of loss function and raise the accuracy of the classification. Besides, as the three classes are data imbalanced, data rebalance is also realized through the cross-entropy loss function by multiplying the proportions of these three classes in the training set.

\begin{figure*}[ht]
	\centering
	\includegraphics[width=1.\linewidth,angle=0]{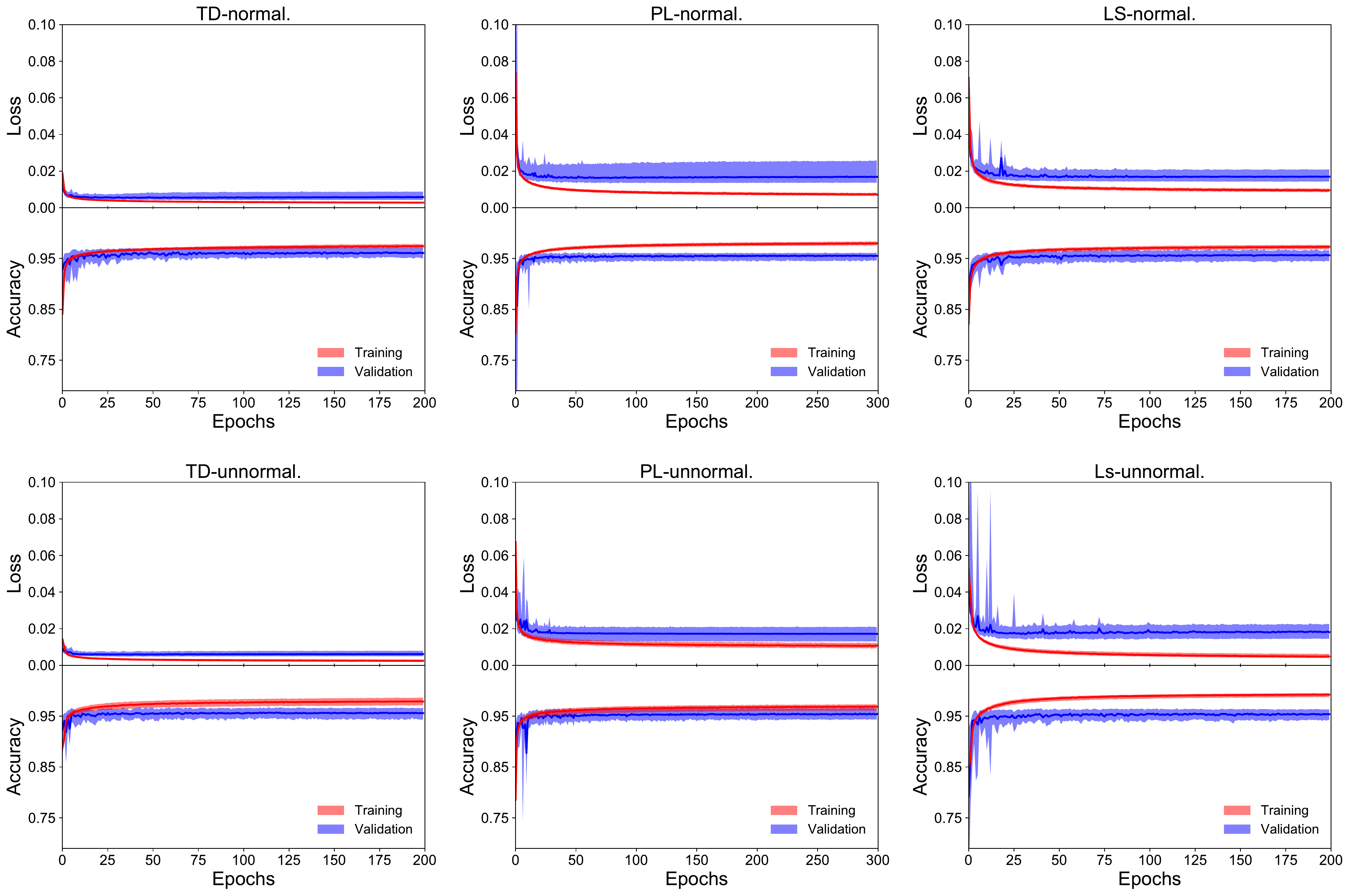}
	\caption{The loss and accuracy vs. training epochs for six CNNs. The outputs of each network give the probabilities of classifying into the three classes. The gold class and silver class both stand for superflare events. The accuracy represents the right-classification rate of superflare and nonsuperflare events. In another words, even if the outputs give three probabilities, we only distinguish superflare and nonsuperflare events. The results are from 70 training and validation data sets. The blue and red represent the results of validation and training set as the training epochs increase. The solid colored lines stand for the median values of results from all data sets. The shaded area stands for 5\% to 95\% of each result.}
	\label{fig:CNN-loss-accu}
\end{figure*}

The results of 70 training and validation data sets are shown in Figure \ref{fig:CNN-loss-accu}. These can be also treated as results of the $k$-folding cross validation. Note that we classify candidates into three classes; however both the gold class and silver class contain superflare events. So, results from these two classes are combined to calculate the accuracy of a network. The reason why we set three classes has been illustrated in Section \ref{sec:superflare candidates}. 
From the figure, we can see that the loss and accuracy of the six networks are more stabilized as the training epochs increase. Some of networks show some overfitting as the accuracy of the validation is relatively lower than that of the training set. 
This is caused by two reasons. First, we only have about 15,000 data, and the networks may be relatively complicated with many neural units, which should be better constrained by even more data. Second, we tend to choose those networks that perform well on the validation data with results of higher accuracy and stable training progress. We have also tried to simplify those networks or put a penalty on loss function with a higher weight decay. However, none of these enhance the performance of networks, but reduce it. So, we control overfitting by adding dropout layers in-between convolutional layers and setting the learning rate decay to avoid relatively higher learning rate as the learning epochs increase.

\begin{equation}\label{equ:confumatrix}
\begin{tabular}{cccc}
                              &                            & \multicolumn{2}{c}{CNN results}                   \\
                              &                            & Positive                    & Negative                   \\ \cline{3-4} 
\multirow{2}{*}{Real truth} & \multicolumn{1}{c|}{Positive}  & \multicolumn{1}{c|}{$TP$} & \multicolumn{1}{c|}{$FN$} \\ \cline{3-4} 
                              & \multicolumn{1}{c|}{Negative} & \multicolumn{1}{c|}{$FP$} & \multicolumn{1}{c|}{$TN$} \\ \cline{3-4} 
\end{tabular}
\end{equation}

In Equation \ref{equ:confumatrix}, we list an example of the confusion matrix, which is helpful when we introduce recall, precision, the true-positive rate (TPR) and false-positive rate (FPR).
In this matrix, TP, FN, FP, and TN stand for numbers of true-positive, false-negative, false-positive, and true-negative events. In this work, true events are superflare events. The real truth is artificial classification. The CNN results are classifications based on CNN outputs. So, the accuracy, recall ($R$), precision($P$), TPR and FPR can be derived as
\begin{equation}\label{equ:accuracy}
	{\rm Accu.}=\frac{TP+TN}{TP+FN+FP+TN},
\end{equation}
\begin{equation}\label{equ:recall}
	R=TPR=\frac{TP}{TP+FN},
\end{equation}
\begin{equation}\label{equ:presicion}
	P=\frac{TP}{TP+FP},
\end{equation}
\begin{equation}\label{equ:FPR}
	FPR=\frac{FP}{FP+TN}.
\end{equation}
The break-Even Point (BEP) and area under the receiver operating characteristic (ROC) curve (AUC) are also derived from those network results.
BEPs are points where precision equals recall in the PR curve, which are derived by setting limits at each of the ranked possibilities of positive events to calculate the specific recall and precision. The ROC curve also sets these limits and calculate the specific TPR and FPR. The AUC is the area under this curve. The best network is with Accu.$=100\%$, $R$=TPR=1, P$=1$ and FPR$=0$, from BEP$=1$ and AUC$=1$, which means that the network can ideally recall all positive events with 100\% precision and perfectly distinguish positive events from negative events.

Each of the seven test data sets is inputted into the corresponding networks. For example, the test set in Figure \ref{fig:CNNdataset} is stack 0, which is inputted into networks trained by using data from a combination of ministacks without ministack 0. Then, the same test set is inputted into networks trained by data without ministack 1, and so on. This test set is inputted into the corresponding networks 10 times, as do other test sets. Finally, we get 70 results for each of the six networks. Their accuracy, recall, and precision are all listed in Table \ref{tab:ResultsCNNs}. As these results are from the test data set, these properties can indicate the generalization ability of the CNN. These results are based on the classification of superflares or nonsuperflares instead of three classes, as we illustrated above. 

\section{Ensemble learning}
\label{sec:ensemble learning}
The six CNNs for the TD arrays, PL images, and LS images do not show perfect performance with even higher accuracy. So, we try to find a way to improve accuracy for classifying superflare candidates. After training the six separate networks, ensemble learning is a handy way, including the stacking and voting methods. Here, we illustrate these two methods, through which we improve the superflare classifying accuracy.

\subsection{Stacking}
\label{sec:stacking}
\begin{figure*}[ht]
	\centering
	\includegraphics[width=0.5\linewidth]{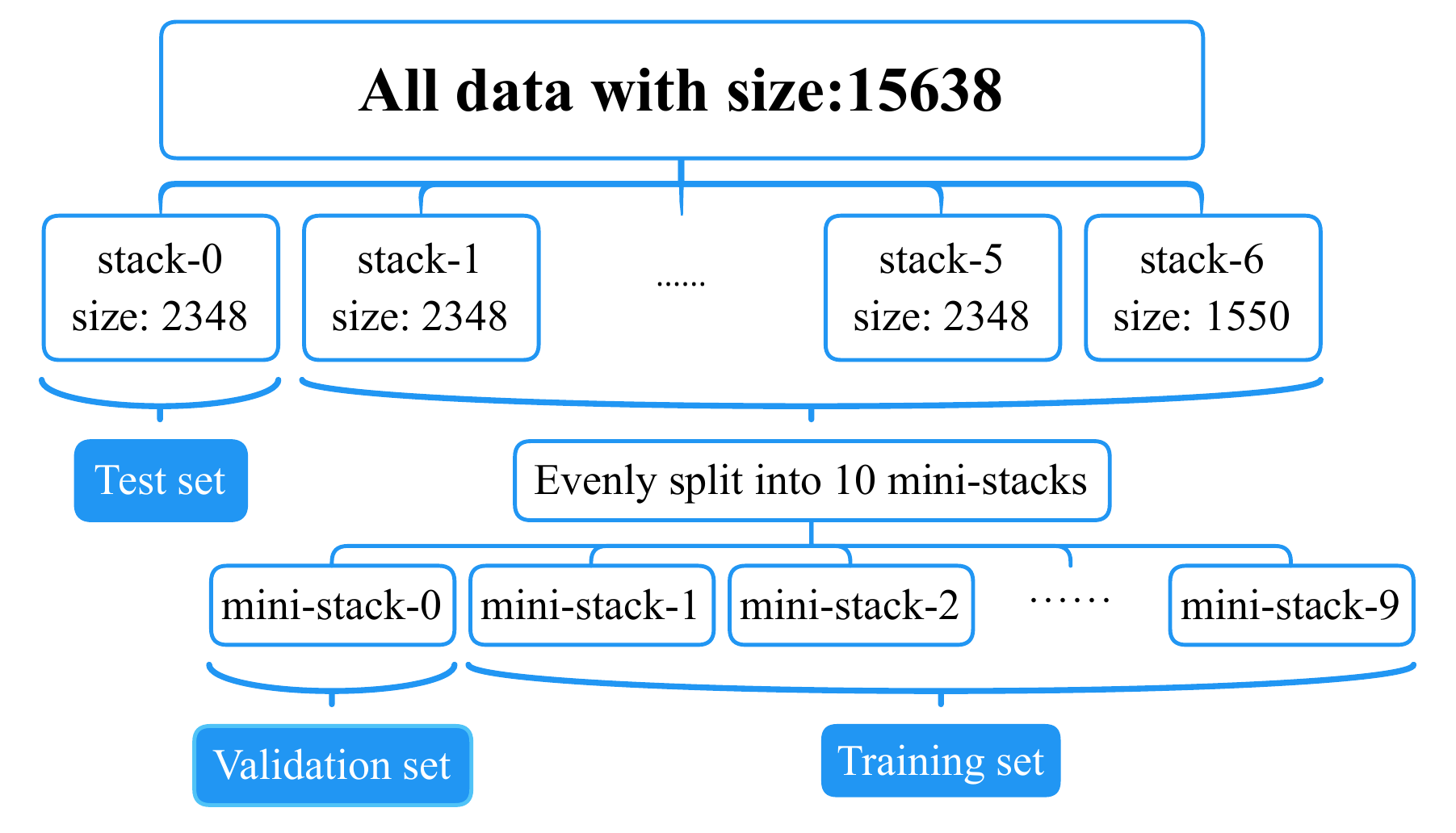}

	\caption{Flow chart for setting the training, validation, and test data sets. As shown by this workflow, we have seven test data sets, each of them corresponds to 10 validation and 10 training sets. The data balance between three classes is considered when we split the data into stacks and ministacks. The training set, validation set, and test set are 75\%, 10\%, and 15\% of the whole database. Separation into stacks and ministacks is for $k$-folding cross validation.}
	\label{fig:CNNdataset}
\end{figure*}

The stacking method can also be understood as training another CNN network (hereafter stacking CNN), whose inputs are a combination of those outputs from many individual networks \citep[][]{WOLPERT1992241,breiman1996stacked}. 
As we introduced in Section \ref{sec:cnn datasets}, we have seven test sets $\times$10 training and validation sets. Each test set, validation set, and training set are independent. For one test set, there are 10 training and validation sets. In order to avoid overfitting of training the stacking CNN, outputs of inputting training data set into trained networks (six CNNs) are not used for training the stacking CNN \citep{zhou2021machine}. These six networks are illustrated in Section \ref{sec:six cnns}. For instance, stack 0 in Figure \ref{fig:CNNdataset} is treated as a test set. First, we input the validation set ministack 0 and the test set into six networks, which are trained by using the remaining ministacks, and we get the first 1/10 training data (train-part-0) and first test set (test stack 0) for the stacking CNN. Then, we use the ministack 1 and the same test set into the networks, which are trained by the remaining ministacks, and get the second 1/10 training data (train-part-1) and second test set (test stack 1), and so on. Finally, for the test set stack 0, we have 10 test stacks and one training data combined by the 10 train parts. There are seven test sets (aka stacks in Figure \ref{fig:CNNdataset}). So we get seven corresponding training sets. Each one of them corresponds to 10 test stacks for stacking CNN, respectively. The outputs of the six networks are combined as an array with size of 6$\times$3 (each of the six networks outputs an array with size 1$\times$3) and inputted into the stacking CNN. The new 7$\times$10 test stacks are used for the $k$-folding cross validation of the stacking CNN. The workflow of the stacking CNN can be found in Figure \ref{fig:stacking-voting-flows}(a). The results of putting 70 test stacks into the trained stacking CNN are listed in Table \ref{tab:ResultsCNNs}. These 70 test stacks are also used in the following voting method.

\subsection{Voting}
\label{sec:voting}
Voting is another method to integrate many separate CNNs, whose outputs are treated as votes to classify the candidates. Here, we use two voting actions including soft voting and hard voting, which can be formulized as 

\begin{equation}
S(x)= [s_{i}(x) \;{\rm for}\; i\in \{0,1,2 \}], 
\end{equation}

\begin{equation}\label{equ:soft-voting}
{\quad \rm where \;} s_{i}(x) = \frac{\sum_{j=1}^{6} c_{i}^{j}(x)} {\sum_{i=0}^{2}\sum_{j=1}^{6} c_{i}^{j}(x)},
\end{equation}

and

\begin{equation}
H(x)= [h_{i}(x)/3 \;{\rm for}\; i\in \{0,1,2 \}],
\end{equation}

\begin{equation}\label{equ:hard-voting}
{\rm where}
h_{i}(x)=\bigg\{
\begin{aligned}
{h_{i}(x)+1\;(\arg\max\;\{c^{j}(x)\}={i}\;{\rm for}\; j\in \{1,2,3,4,5,6 \})} \\
{h_{i}(x)+0\;(\arg\max\;\{c^{j}(x)\}\ne{i}\;{\rm for}\; j\in \{1,2,3,4,5,6 \})}
\end{aligned}
.
\end{equation}
Here, $S(x)$ and $H(x)$ stand for soft voting and hard voting, and $x$ is the candidate data needed to be classified. $c_{i}^{j}$ is the output applied by theSoftmax function in Equation \ref{equ:softmax}. $i$ is the $i^{\rm th}$ class, as we classify candidates into three classes (gold class, silver class and none class), $i\in \{0,1,2 \}$. $j$ is the $j^{\rm th}$ network from 1 to 6. The corresponding orders of networks are listed as the first six rows of Table \ref{tab:ResultsCNNs}. The specific workflows of voting are shown in panel (b) and (c) of Figure \ref{fig:stacking-voting-flows}. In order to strictly classify candidates through the voting method, we set a limit at 75\%. So, $\max H(x)$ and $\max S(x)$ should be larger than this limit; then the candidate can be labeled. Otherwise, the voting method refuses to label the candidate, and the data remain for artificial labeling after visual inspection. For hard-voting, $75\%$ means more than four out of six individual networks have the same voting result. As some candidates are denied classification, we define the classification rate as 
\begin{equation}\label{equ:classrate}
	{\rm Class. rate}=\frac{N-N_{\rm TBD}}{N},
\end{equation}
where $N$ is the data count of all data of a test stack, and $N_{\rm TBD}$ is the number of data that are cannot be classified and are labeled as `TBD' in the voting flows.
As the outputs derived from the Softmax function can be treated as possibilities for classifying a candidate, we regard that each of the six CNNs has 100 votes. For soft voting, $75\%$ means that more than $75\%$ of 600 votes support the same voting result. We do not set this kind of limit for those six separated CNNs. The reasons are as follows. (1) There is no practical meaning for the limit, if we set it for each of the six CNNs. (2) It is hard to get a specific percentile limit that is reasonable and explicable. (3) Even if we set $75\%$ as limits for each of the six CNNs, accuracy and other properties do not achieve better results than voting methods. (4) Voting from the six CNNs can deny to classify those candidates that may be hard to classify or have not been used previously to train the CNNs. So, we insist to use the maximum outputs of each of the six CNNs to classify candidates instead of set a limit. The results of the voting methods are listed in Table \ref{tab:ResultsCNNs}.

\section{Results and discussions}
\label{sec:results and discussions}

\subsection{Public open data set}
\label{sec:public open data set}
In Section \ref{sec:superflare candidates}, we have introduced the method through which we select superflare candidates from three years of {TESS} observations. These 15,638 flare candidates are all from solar-type stars (with stellar effective temperatures of $5100 \mathrm{K} \leqslant
T_{\mathrm{eff}}<6000 \mathrm{K}$ and surface gravities of $\log g >4.0$), and their information are listed in Table \ref{tab:candidates}. It is worth noting that only 5060 among all 15,638 candidates are superflare events, and others are nonsuperflare data. The corresponding stellar properties of their target stars are listed in Table \ref{tab:solar type stars}. Note that, although the stellar periods may be a combination of three years observations, we should be careful when using periods larger than 10 days because a sector of {TESS} has a limited observing span of about 27 days, and the observation of each sector is normalized by itself \citep{2020ApJ...890...46T}. We classify these data into three classes: gold class, silver class, and none class, among which the gold and silver classes collect superflare events. As these three classes have different specialties, and in order to improve the performance of CNNs, we still use three classes for training the CNNs, but use superflare events or nonsuperflare events as a binary classification problem to calculate the properties of the CNN. The three classes with their own characteristics are articulated in Section \ref{sec:features_pics} and shown in Figure \ref{fig:GoldSilverNone}. 

We emphasize that these candidates are classified into three classes.
However, we do not statistically use these data to do any research like \cite{2020ApJ...890...46T, 2021ApJS..253...35T}, which is beyond the topic of this article. Using the CNNs in this work to process the four years of {TESS} observations will accomplish collecting all superflare events detected by {TESS}. Following that, statistical research can be performed, for instance, comparing stellar flare frequencies on different types of star and comparing stellar flares on one type of stars but with different ages.
For readers who are interested into these data, they can find corresponding events as specific information are released. One could also use these data to get creative feature images or data for their own CNN training. Besides, the six kinds of data sets including TD-normal., TD-unnormal., PL-normal., PL-unnormal., LS-normal., and LS-unnormal. are all public released \footnote{\url{https://github.com/TuinGIt/CNNforTESS}}. Their corresponding file names are also listed in Table \ref{tab:candidates}. So, these data sets are really useful and valuable for the community of searching for stellar superflares. They are useful for searching for superflares on other types of stars, as deciding the type of a star is totally irrelevant with detecting superflare candidates, which is only according to light-curve and pixel-level data. 

\subsection{Comparing CNNs}
\label{sec:comparing cnns}
We compare results of these networks and methods listed in Section \ref{sec:cnn datasets}. These result are listed in Table \ref{tab:ResultsCNNs}. The classification rate, accuracy, recall, precision, BEP and AUC are used to indicate the generalization capabilities of different CNNs and methods. Note that: (1) In order to compare results from those six different CNNs, we use the same seven test sets (stacks in Figure \ref{fig:CNNdataset}). Each one of these six CNNs is trained once by each of 10 training data sets, so we get a total of 70 outputs (7$\times$10). The only difference is the inputs of these CNNs, which are different corresponding feature images or arrays. For the voting and stacking methods, 70 test stacks are inputted into the corresponding stacking CNNs, which are trained by each of the seven training sets. The processes of setting the data sets for deep learning and ensemble learning are introduced in Section \ref{sec:CNNs} and Section \ref{sec:ensemble learning}. So, the original candidates and counts of data in each of three classes are all the same when comparing these CNNs and voting methods. (2) The ultimate purpose of this work is to efficiently distinguish superflare events and nonsuperflare events. Although we set three classes to improve the performance of the six CNNs, the results according to the two-class classification (superflare or nonsuperflare) are obtained. According to the results of the two-class classifications, we discuss using these CNNs to search for superflares in the future.

\begin{figure*}[ht]
	\centering
	\includegraphics[width=0.5\linewidth,angle=0]{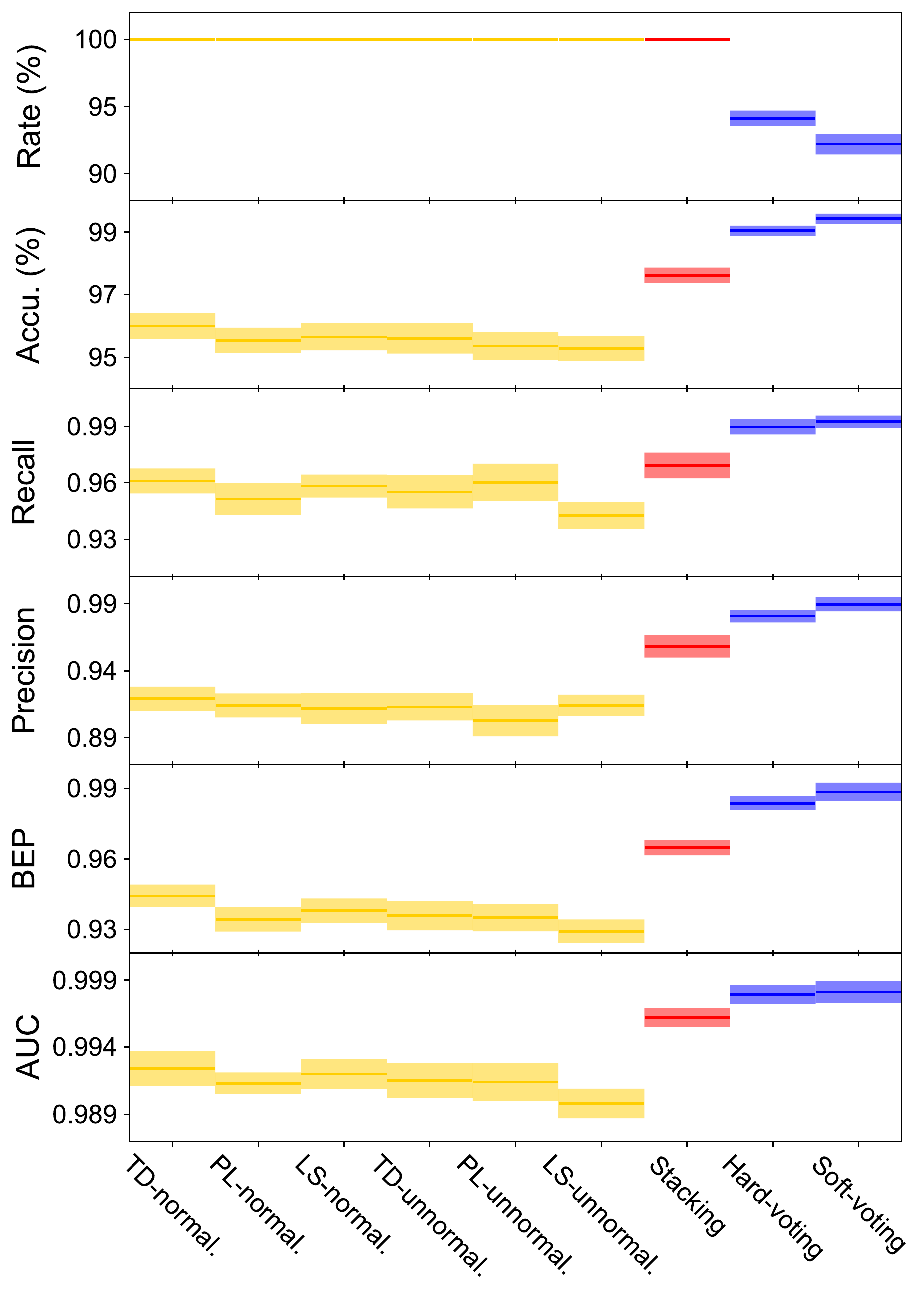}
	\caption{Visualization of properties in Table \ref{tab:ResultsCNNs}. The solid lines and shaded area show the mean values and standard deviations of each property. Yellow, red and blue stand for the six CNNs, stacking CNN, and voting methods, respectively.}
	\label{fig:comparing}
\end{figure*}

\begin{figure*}[ht]
	\centering
	\includegraphics[width=1\linewidth,angle=0]{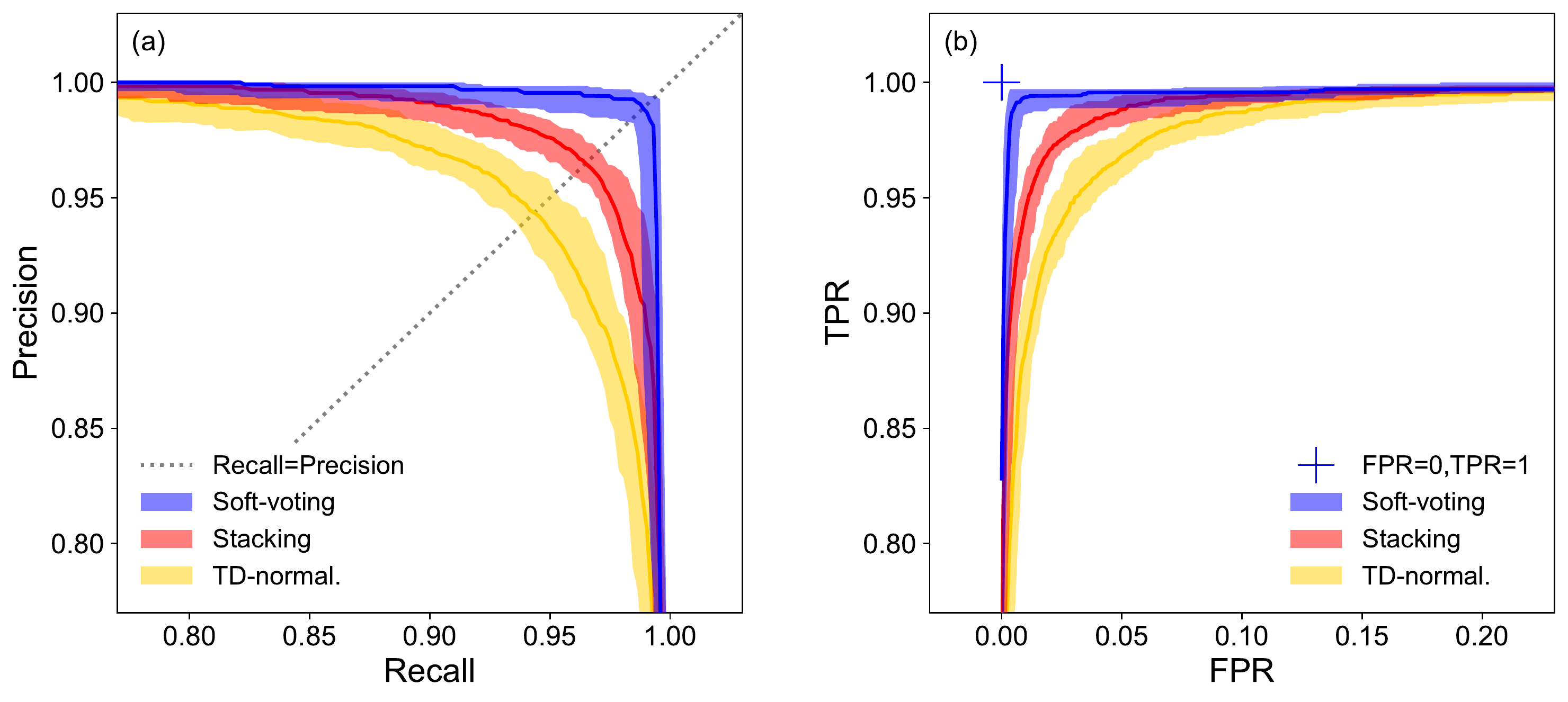}
	\caption{The PR curve and ROC curve are shown in panels (a) and (b). The dashed gray line in panel (a) represents ${\rm Recall}={\rm Precision}$. The blue cross mark in panel (b) represents the point where ${\rm FPR}=0,\; {\rm TPR}=1$, This is the result from the best ideal network, as all positive candidates are selected without any negative ones. As TD-normal. is the best network of the six CNNs and soft voting is better than hard-voting method, we use the results from these two to compare with the stacking CNN. The results from soft voting, stacking CNN, and TD-normal. are shown in blue, red, and yellow. The solid curves represent the median values of the results. The shaded area denotes the ranges of each property from 5\% to 95\%.}
	\label{fig:PRROC}
\end{figure*}

For comparing these CNNs and voting methods, the properties in Table \ref{tab:ResultsCNNs} are visualized in Figure \ref{fig:comparing}. From this figure, the accuracy of these six CNNs is around 95.5\%.  
It is obvious that TD-normal. is the best network of these six CNNs, and we use the properties of it to compare with stacking and soft voting in Figure \ref{fig:PRROC}. 
As the shaded area (std intervals) of the six CNNs is relatively larger than that of the stacking and voting methods, the six CNNs may be not very stable compared to the stacking CNN and voting methods according to all test sets.
We also compare PR curves and ROC curves of the TD-normal. stacking CNN and soft voting in Figure \ref{fig:PRROC}. Both of these curves indicate the generalization and performance of a CNN. 
Comparing with the CNN of TD-normal., 
the accuracy and other properties of stacking CNN are significantly enhanced. 
The six CNNs and stacking CNN are both classifying all candidates with a 100\% classification rate, unlike the voting methods. This notable enhancement strengthens the idea that each of six CNNs and inputs of them have their own individuality. A combination of these six CNNs could improve redundancy or error-tolerant rate of searching for superflares. From Figure \ref{fig:PRROC}, it is noticeable that the performances of the ensemble-learning methods are more enhanced than those of the six CNNs. Only through one CNN obtaining a higher generalization performance is not that practical as we only use less than 15,368 data to train a network, and only about 1/3 of them are labeled as superflare events. The well-known deep-learning benchmark database MINIST \citep{726791,KUSSUL2004971} and Fashion-MNIST \citep{2017arXiv170807747X} have 60,000 training instances, even if they are classified into ten classes, and each class is definitive with clear boundaries. 
In this work, only superflare events are distinguished from other nonsuperflare events with clear features and boundaries. However, it is also possible to classify many subclasses in the none class, like exoplanet transients, cosmic ray events, CCD errors, and pipeline noise. 
For convenience, we blend all these nonsuperflare events in the none class to distinguish from superflare events, which may confuse CNNs when nonsuperflare events are inputted. So, limited instances of superflares and blending all nonsuperflare events together may be the reason why we do not get a single CNN, which is better, more stable and comparable to ensemble-learning methods yet. Using ensemble-learning methods is more complicated as we have to train and get six kinds of CNNs and input data, but ensemble-learning is now a practical and relatively powerful method when we try to classify superflare candidates. Besides, the stacking CNN, hard voting and soft voting are more stable and better than six CNNs according to Figure \ref{fig:comparing}. 

\begin{figure*}[ht]
	\centering
	\includegraphics[width=1.\linewidth,angle=0]{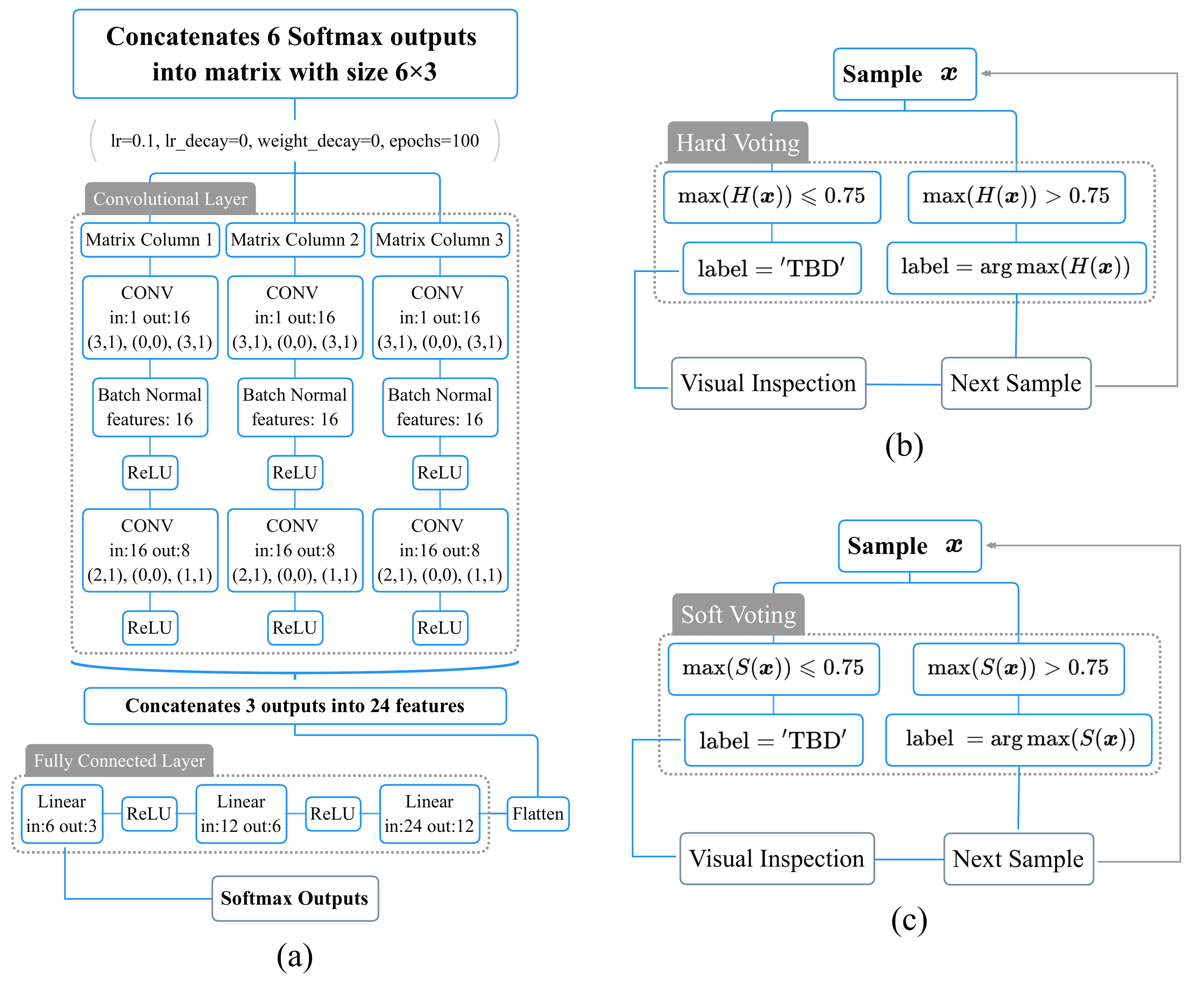}
	\caption{The workflows of the stacking CNN, hard voting and soft voting are shown in panel (a), (b) and (c), respectively. The inputs of the stacking CNN are arrays with size 6$\times$3, and the values coming from the combined outputs of the six CNNs are shown in Figure \ref{fig:CNN-flow}. In the voting flows, ${\rm label} = '{\rm TBD}'$ means that the classification of the candidate is to be determined after visual inspection.}
	\label{fig:stacking-voting-flows}
\end{figure*}

If we describe the enhancement of stacking CNN as taking advantage of the differences between the six CNNs and their inputs, the voting methods are more emphasized to use the result consistency of these six CNNs. For the voting methods, the accuracy, recall, precision, BEP and AUC are calculated based on those classified candidates apart from others labeled as `TBD' after voting, as shown in Figure \ref{fig:stacking-voting-flows}(b) and \ref{fig:stacking-voting-flows}(c). 
Comparing the stacking CNN with the voting methods in Figure \ref{fig:comparing}, it is evident that the voting methods are more stable and exceptional, and soft voting performs better than hard voting. 
In Figure \ref{fig:PRROC}, soft voting is really close to the ideal vision of a classification machine-learning method with $R\sim1$, $P\sim1$, FPR$\sim 0$ and TPR$\sim 1$. However, the classification rate decreases from 100\% to about 92\%, as shown in Figure \ref{fig:comparing}. 
This is a trade-off between performance and the classification rate of a CNN. The voting methods evenly consider each of the six CNNs, through uniting outputs of these six CNNs. As we set 75\% as a limit over which a candidate is classified, this can be understood to ensure voting results are broad, representative and accurate among these six CNNs. 
We also checked those candidates whose maximum voting results were not up to a 75\% limitation. These candidates are intractable, even through visual inspection of their signal-to-noise ratios (S/N) are relative low, and hard to be classified. This may result from the fact that each of the six CNNs is also hard to be trained to detect superflare features at such low S/N. 
Another reason might be that some kinds of data have not been paid adequate attention to as their data are not enough in the training sets. These kinds of data may not be well learned by CNNs and rejected for classification. The voting method maximally exposes these unclassified data with an about 8\% exposure rate ($100\%-92\%$) for soft voting. Besides the decrease in the classification rate (derived in Equation \ref{equ:classrate}), other indicators of network generalization and performance are all enhanced distinctly, as shown in Figure \ref{fig:comparing}, and accuracy is improved to values even higher than 99\%. So, we could improve the accuracy of soft voting to more than 99.42\% assuming that the manual visual inspection is 100\% accurate and trustworthy. 
As some kinds of data are not enough to be paid adequate attention to while training CNNs, it is likely that some data may be new transient phenomena. Through one-on-one specifically visual inspection, we will find these new phenomena instead of submerging them into endless data. 
Back to our initial purpose of using CNNs for searching superflare events, we pursue an ideal vision that those automatically searched events are all true-positive superflare events in a database. At the present moment, the benefits of artificial intelligence are more reflected in efficiency while processing some repeatable jobs instead of handling extraordinary unseen situations. We still recommend readers who are willing to use our CNNs for searching superflare events to at least visually inspect the remaining not classified 8\% data. These data can also be used for enlarging the superflare database for training the six CNNs or other networks in the future. 

\subsection{Duration and amplitude}\label{sec:DuraAmp}
\begin{figure*}[ht]
	\centering
	\includegraphics[width=1\linewidth,angle=0]{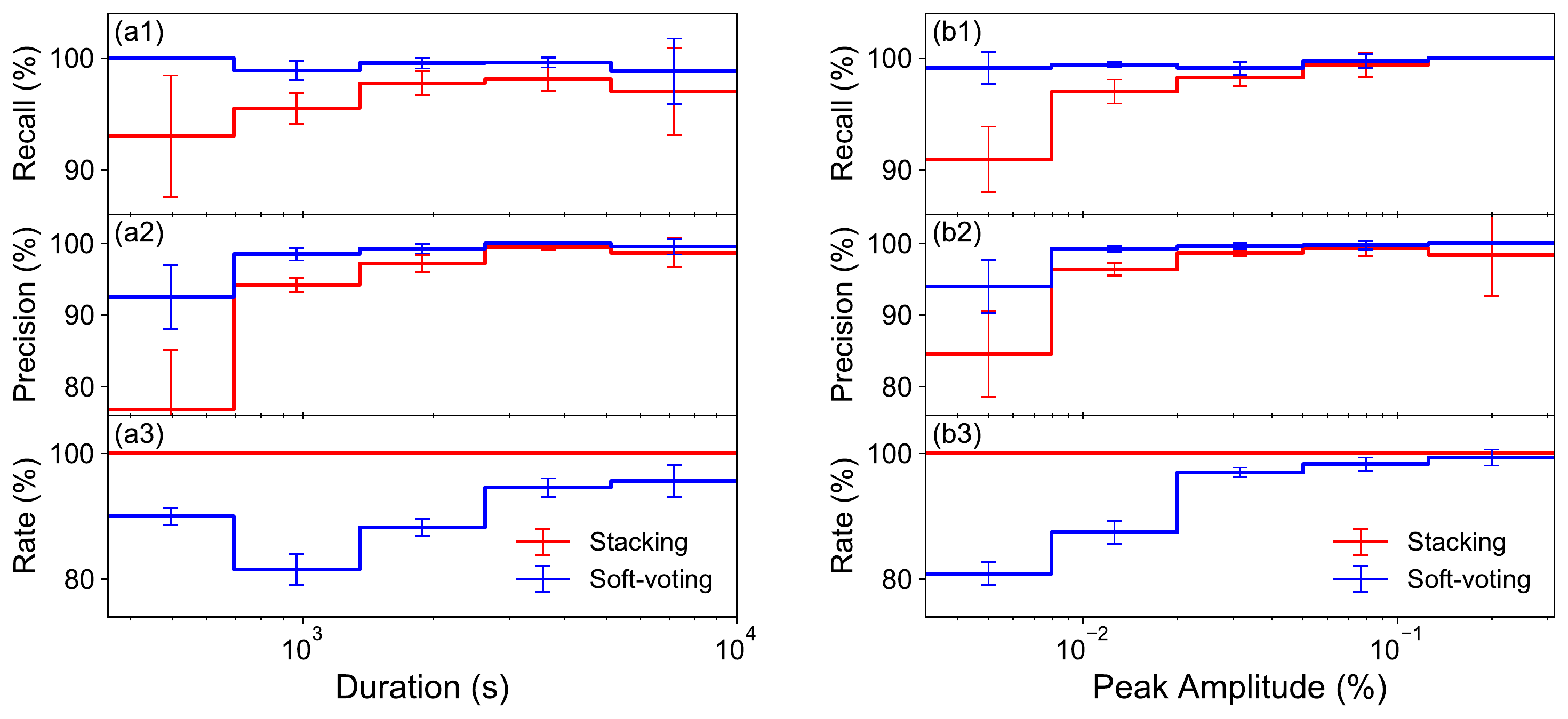}
	\caption{Specific recall, precision and classification rate in each bin of superflare duration and amplitude. The bin size of the duration and amplitude is in logarithmic scale. We compare the stacking CNN with soft voting, which are represented by red and blue curves, respectively. Recall and precision are defined in Equations (\ref{equ:recall}) and (\ref{equ:presicion}). Rate stands for classification rate defined in Equation (\ref{equ:classrate}). These three properties are all in unit of percentile in this figure. The error bars are standard deviations of each property. The data from which we plot this figure are the same as those of Figure \ref{fig:comparing}, Figure \ref{fig:PRROC} and Table \ref{tab:ResultsCNNs}, but with specific results from the bins of duration and peak amplitude.}
	\label{fig:DuraAmpl}
\end{figure*}

As we have discussed, superflare candidates with relatively low S/N are hard to be visually inspected. The same also applies to those six CNNs trained before. 
It is possible that the duration and amplitude of a superflare can be related to the S/N of a superflare.
These two properties can be intuitively recognized. 
Those superflares with long durations and large amplitudes are more obvious and have abundant superflare features. In Figure \ref{fig:DuraAmpl}, it is obvious that the performances of the stacking CNN is not better than that of soft voting, as shown in panels (a1), (a2), (b1) and (b2). The candidates with short durations and low peak amplitudes are not better classified than these with long durations and high amplitudes. In another words, the six trained CNNs or stacking CNN do not perform well on candidates with relatively lower S/N. For soft voting, greater performance is achieved by rejecting the classification of candidates with shorter durations and lower peak amplitudes, as shown in panels (a3) and (b3) of Figure \ref{fig:DuraAmpl}. So, when using the classification results from the CNNs in this work, we should pay more attention to those candidates that are classified as superflare events but with short durations and low peak amplitudes. In the future, the updates of the six CNNs and stacking CNN presented in this work or CNNs that may be constructed should lay emphasis on superflare instances with lower S/N. 
Here, we propose two ways: (1) collect more data with relatively lower S/N, and (2) make the time interval of a superflare at the same scale of width for TD arrays, PL images and LS images, through stretching or compressing the time axis.

\section{Summary}
\label{sec:Summary}
This work focuses on developing CNNs for superflare automatic searching on pixel-level data. We use 15,638 superflare candidates, which are classified into three categories, namely, the gold class, silver class and none class. Examples of these classes are shown in Figure \ref{fig:GoldSilverNone}. These candidates are from 71,732 solar-type stars (with $5100 \mathrm{K} \leqslant
T_{\mathrm{eff}}<6000 \mathrm{K}$; $\log g >4.0$), which are observed by three years of {TESS} observations. These candidates are automatically searched through our previous methods \citep[][]{2021ApJS..253...35T,2020ApJ...890...46T} and visually inspected in this work.
The information of solar-type stars and superflare properties are listed in Table \ref{tab:solar type stars} and \ref{tab:candidates}.

Six CNNs have been trained and validated by $k$-folding cross validation. These six CNNs are trained by feature data including TD arrays, PL images, and LS images. All superflare candidates and feature data are public available. Examples of these feature data are visualized in Figure \ref{fig:feature_pics}. As they are uneven counts of instances in the gold class, silver class, and none class, data rebalancing is considered when training the CNNs. These six CNNs are different from each other and are developed considering the characteristics of different features.

Ensemble learning is used when combining all outputs of the six CNNs. The six trained CNNs, stacking CNN, and code are publicly available \footnote{\url{https://github.com/TuinGIt/CNNforTESS}}. After comparing indicators of the networks' performance, like the accuracy, recall, precision, BEP, and AUC of the CNN, we found that the stacking and voting methods evidently enhance the performance and generalization ability of the six CNNs. The reason is that the combination of the six CNNs depicts superflare features according to abundant points of view much more than a single CNN. The stacking CNN improves the redundancy or error-tolerant rate of searching superflares. Voting methods including soft voting and hard voting show better performance than the stacking CNN, but the trade-off classification rate decreases from 100\% to about 92\%. Meanwhile, the accuracy is improved from 97.62\% to over 99\%. The voting methods achieve better result consistency than the six CNNs. We set a limit of 75\%, which ensures that votes united from the six CNNs are broadly representative, more accurate, and reliable while searching for superflares. Soft voting is the best method with $R\sim1$, $P\sim1$, FPR$\sim 0$ and TPR$\sim 1$. 8\% of unclassified data visually inspected one by one. These data are basically intractable with relatively low S/N and hard to be classified. Visual inspection of these data will not only improve the accuracy to higher than 99.42\%, but also enlarge the size of the training database for updates of the six CNNs updates in the future.

The performance of these six CNNs is better on superflare candidates with high S/N (also with longer durations and higher peak amplitudes). The voting methods are more capable of rejecting those candidates with shorter durations and lower peak amplitudes. This is the reason why soft voting is better than stacking CNN. The voting methods also give chances for finding new transient phenomena, which are not recognized by trained CNNs at all and will not be successively classified but left for visual inspection.

In this work, six CNNs and ensemble-learning methods are developed for the purpose of automatically identifying superflare events from pixel-level data, which ensures that the selected superflares are true-positive events. A contamination of false-positive events will influence the statistical analysis of superflare data. 
For instance, the number of nonsuperflares events is 2 times more than that of superflare events in the database of this work, whose statistical weight cannot be neglected. Although superflare events in this work are selected from solar-type stars, the training CNNs are only based on observational data. Stellar parameters are not considered. 
So, these trained CNNs and methods can also be applied to the detection of stellar flares of other types of stars. 
CNN training is time and data consuming. Performance and generalization of a CNN is also based on the scale of database. In the future, updating CNNs of this work should be still based on larger scale of superflare database, especially superflares with low S/N. CNNs will efficiently process archived and upcoming data from {TESS}. 
\citet{2021NatAs.tmp..246N} probably found a stellar filament eruption associated with a superflare event through optical spectroscopic observation. If CNN forecasting of stellar superflare is available in the future, it will buy time for preparing follow up optical spectroscopic observation. If so, the work by \citet{2021ApJ...916...92W} could also capture the whole spectroscopic observations from before until the end of the stellar flare, and \citet{2021NatAs.tmp..246N} would not spend too much precious ground observation time on a single star. They would not probably find stellar filament eruption only by chance. 
Much more stellar coronal mass ejection events could be more efficiently detected. We can foresee that the scale of stellar superflares can be efficiently enlarged by CNNs in this work and more interesting statistical research could be done in the future, including but not limited to comparison of superflares on different types of stars. 

\section*{Acknowledgements}
We thank two anonymous referees for helpful comments. We would like to thank B. B. Zhang, Niu Liu and Bo-yang Wang for valuable discussions and helps. We sincerely thank the Mikulski Archive for Space Telescopes (MAST) and the {TESS} community for applying reachable data portal and tools. This work is supported by the National Natural Science Foundation of China (grant No. U1831207), the Fundamental Research Funds for the Central Universities (No. 0201-14380045), and the China Manned Spaced Project (CMS-CSST-2021-A12). The codes of this work were run on a high-performance computing equipped with eight NVIDIA Tesla P100 GPU computing processors \footnote{https://www.nvidia.com/en-sg/data-center/tesla-p100/}. 

\facilities{TESS, Gaia, Hipparocos.}

\software{Jupyter Notebook \citep{Kluyver2016jupyter}, numpy \citep{2011CSE....13b..22V}, matplotlib \citep{2007CSE.....9...90H}, pandas \citep{mckinney-proc-scipy-2010}, lightkurve \citep{2018ascl.soft12013L}, astropy \citep{2013A&A...558A..33A}, astroquery \citep{2019AJ....157...98G}, pytorch \citep{NEURIPS2019_9015}.}

\bibliography{export-bibtex}{}
\bibliographystyle{aasjournal}

\FloatBarrier
\begin{table*}[!htb]
    \renewcommand{\arraystretch}{1}
	\addtolength{\tabcolsep}{+10pt}
	\caption{Properties of solar-type stars.}
	\label{tab:solar type stars}
	\centering
    \begin{tabular}{lccccccc}
        \hline
        \hline
{TESS} ID & ${\rm T}_{\rm mag}$ $^a$& $T_{\rm eff}$ $^b$ & $\log{g}$ $^c$ & Radius $^d$    & Period  $^e$ & Rvar  $^f$   & Flag $^g$\\
                                   &                     & ($K$)         &           & ($R_{\odot}$) & ($days$) &          &      \\ \hline
260268381                          & 9.22                & 5894          & 4.00      & 1.71          & 5.73     & 0.002219 & ···  \\
152937588                          & 9.12                & 5660          & 4.12      & 1.46          & 3.49     & 0.002008 & ···  \\
410214986                          & 7.85                & 5414          & 4.45      & 0.95          & 3.58     & 0.056958 & HB   \\
260188537                          & 9.31                & 5540          & 4.02      & 1.59          & 7.48     & 0.008919 & GB21 \\
197959959                          & 9.37                & 5353          & 4.36      & 1.05          & 6.97     & 0.005462 & GB42 \\
261463150                          & 10.15               & 5838          & 4.58      & 0.87          & 9.29     & 0.003536 & GM21 \\
141425702                          & 10.99               & 5718          & 4.52      & 0.92          & 13.42    & 0.006033 & GM42 \\ \hline
    \end{tabular}
\begin{flushleft}
		\textsc{Notes.}\\ {
			$^a$ {TESS} magnitude. \\
			$^b$ Surface effective temperature of the star in unit of K. \\
			$^c$ Surface gravity of the star in log scale.\\
			$^d$ Stellar radius in unit of $R_{\odot}$. The above properties are all from {TESS} input catalogue \citep[TIC v8;][]{2019AJ....158..138S}.\\
			$^e$ Periodicity of the star in unit of days.\\
			$^f$ Stellar photometric variability ($R_{\rm var}$) as defined in Equation \ref{equ:rvar}.\\
			$^g$ Flags for stars after crossmatching with the {Hipparocos}-2 and {Gaia} EDR3 catalogs. Here, HB stands for stars of the {Hipparocos}-2 catalog that may be binary systems. GM and GB stand for stars that may have M-type stars or other brighter stars nearby, which are collected by {Gaia} EDR3. 21 and 42 stand for the possible contamination, which are located within 21$''$ or within 21$''$ to 42 $''$ distance from the main targets.\\
			(This table is available in its entirety in machine-readable form.)}\\
	\end{flushleft}
\end{table*}

\begin{sidewaystable*}
    
    \renewcommand{\arraystretch}{1}
    \addtolength{\tabcolsep}{-2pt}
    \caption{Information of superflare candidates}
    \label{tab:candidates}
	\centering
    \begin{tabular}{lrclccccrl}
    \hline
\multicolumn{1}{c}{{TESS} ID} & \multicolumn{1}{c}{Sector $^{a}$} & Source $^{b}$ & \multicolumn{1}{c}{Peak No.$^{c}$} & Label $^{d}$& Peak Date $^{e}$& Peak Lumi. $^{f}$    & Energy $^{g}$  & \multicolumn{1}{c}{Duration $^{h}$} & \multicolumn{1}{c}{File name $^{i}$}                                                                     \\
\multicolumn{1}{c}{}                                   & \multicolumn{1}{c}{}           &        & \multicolumn{1}{c}{}         &       &           & (erg s$^{-1}$) & (erg)    & \multicolumn{1}{c}{(s)}      & \multicolumn{1}{c}{}                                                                              \\ \hline
\multicolumn{10}{l}{Gold-class counts: 1,268}                                                                                                                                                                                                                                                                      \\ \hline
382575967                                              & 7                              & long   & Peak11                       & 0     & 1512.1703 & 1.68E+32       & 4.74E+35 & 6840.11                      & {\tt\string ls\_long\_normalize\_TIC382575967\_Q7\_Peak11.png···}  \\
55752857                                               & 22                             & short  & Peak0                        & 0     & 1923.5225 & 4.39E+32       & 1.07E+36 & 5999.85                      & {\tt\string ls\_short\_normalize\_TIC55752857\_Q22\_Peak0.png···}  \\
260268898                                              & 31                             & short  & Peak0                        & 0     & 2163.4045 & 7.32E+31       & 7.05E+34 & 1920.03                      & {\tt\string ls\_short\_normalize\_TIC260268898\_Q31\_Peak0.png···} \\
260268898                                              & 32                             & short  & Peak0                        & 0     & 2177.9102 & 5.17E+31       & 8.02E+34 & 2880.04                      & {\tt\string ls\_short\_normalize\_TIC260268898\_Q32\_Peak0.png···} \\
260268898                                              & 35                             & short  & Peak1                        & 0     & 2278.9417 & 7.55E+31       & 9.97E+34 & 4079.91                      & {\tt\string ls\_short\_normalize\_TIC260268898\_Q35\_Peak1.png···} \\
\multicolumn{10}{c}{······}                                                                                                                                                                                                                                                                                        \\ \hline
\multicolumn{10}{l}{Silver-class counts: 3,792}                                                                                                                                                                                                                                                                    \\ \hline
339668420                                              & 30                             & short  & Peak5                        & 1     & 2135.0738 & 3.29E+31       & 3.54E+34 & 2040.04                      & {\tt\string ls\_short\_normalize\_TIC339668420\_Q30\_Peak5.png···} \\
235930066                                              & 24                             & long   & Peak1                        & 1     & 1958.7945 & 1.45E+31       & 7.60E+33 & 839.99                       & {\tt\string ls\_long\_normalize\_TIC235930066\_Q24\_Peak1.png···}  \\
294098955                                              & 6                              & short  & Peak0                        & 1     & 1476.7159 & 4.57E+31       & 2.62E+34 & 840.01                       & {\tt\string ls\_short\_normalize\_TIC294098955\_Q6\_Peak0.png···}  \\
339668420                                              & 34                             & long   & Peak0                        & 1     & 2230.0340 & 1.50E+31       & 2.80E+34 & 2640.04                      & {\tt\string ls\_long\_normalize\_TIC339668420\_Q34\_Peak0.png···}  \\
140891597                                              & 28                             & short  & Peak0                        & 1     & 2064.9989 & 7.57E+31       & 1.65E+35 & 3479.99                      & {\tt\string ls\_short\_normalize\_TIC140891597\_Q28\_Peak0.png···} \\
\multicolumn{10}{c}{······}                                                                                                                                                                                                                                                                                        \\ \hline
\multicolumn{10}{l}{None-class counts: 10,578}                                                                                                                                                                                                                                                                     \\ \hline
36828969                                               & 4                              & short  & Peak2                        & 2     & 1423.4314 & -              & -        & -                            & {\tt\string ls\_short\_normalize\_TIC36828969\_Q4\_Peak2.png···}   \\
1119096                                                & 4                              & short  & Peak0                        & 2     & 1423.2551 & -              & -        & -                            & {\tt\string ls\_short\_normalize\_TIC1119096\_Q4\_Peak0.png···}    \\
1119096                                                & 4                              & short  & Peak1                        & 2     & 1423.2912 & -              & -        & -                            & {\tt\string ls\_short\_normalize\_TIC1119096\_Q4\_Peak1.png···}    \\
406413428                                              & 31                             & short  & Peak2                        & 2     & 2161.8467 & -              & -        & -                            &{\tt\string ls\_short\_normalize\_TIC406413428\_Q31\_Peak2.png···} \\
363225381                                              & 39                             & short  & Peak0                        & 2     & 2381.7991 & -              & -        & -                            & {\tt\string ls\_short\_normalize\_TIC363225381\_Q39\_Peak0.png···} \\
\multicolumn{10}{c}{······}                                                                                                                                                                                                                                                                                        \\ \hline
\end{tabular}
\begin{flushleft}
		\textsc{Notes.}\\ {
			$^a$ Sector No. of {TESS}, in which superflare candidates are captured.  \\
			$^b$ `long' and `short' stand for $n=5$ and $n=2$ in Equation \ref{equ:deltaF}. \\
			$^c$ The serial number of a flare candidate captured from the star in a specific sector. For example, the first row stand for the candidate, which is the eleventh one detected from the light curve of TIC382575967 in sector seven, and $n=5$ is set in Equation \ref{equ:deltaF}. \\
			$^d$ Labels of the gold class, silver class, and none class, which are represented by 0, 1, and 2, respectively. All labels are set by visual inspection of the candidates. Here, we have 1268 candidates in the gold class, 3792 candidates in the silver class, and 10,578 candidates are in the none class.\\
			$^e$ Superflare peak time stamps of { TESS}.\\
			$^f$ Superflare peak luminosity in unit of erg s$^{-1}$.\\
			$^g$ Superflare energy in unit of erg, which is calculated by Equation \ref{equ:flareenergy}.\\
			$^h$ Superflare duration in unit of $s$, which is calculated by $t_{\rm end}-t_{\rm start}$. The start and end times of superflare are derived through Equation \ref{equ:errflux}.\\
			$^i$ File name of feature images and arrays of the candidate. These files are used for training CNNs. Their names are structured by types of data, the number of sector, source, and the serial number of the candidate. Note that, data of TD-normal. and TD-unnormal. are not saved in any picture formats but in array format of {\tt\string Numpy} (a {\tt\string Python} package) with file name extension {\tt\string .npy}.  \\
			(This table is available in its entirety in machine-readable form.)}\\
	\end{flushleft}
\end{sidewaystable*}


\begin{table*}
\renewcommand{\arraystretch}{1}
    \addtolength{\tabcolsep}{-1pt}
    \caption{Results of CNNs}
    \label{tab:ResultsCNNs}
	\centering
	
\begin{tabular}{clcccccc}
\hline
\hline
& \multicolumn{1}{l}{Network} & Class. Rate & Accuracy        & Recall              & Precision           & BEP                 & AUC                 \\ 
&  & (\%) & (\%)        &               &            &                  &    
        \\
\hline
\multirow{6}{*}{\shortstack{Six\\CNNs}}                                                    & TD-normal.                  & \textbf{100} & \textbf{96.00 $\pm$ 0.41} & \textbf{0.9609 $\pm$ 0.0066} & \textbf{0.9193 $\pm$ 0.0090} & \textbf{0.9442 $\pm$ 0.0048} & \textbf{0.9924 $\pm$ 0.0013} \\
                                                                             & PL-normal.                  & 100 & 95.54 $\pm$ 0.40 & 0.9514 $\pm$ 0.0085 & 0.9144 $\pm$ 0.0089 & 0.9343 $\pm$ 0.0052 & 0.9913 $\pm$ 0.0008 \\
                                                                             & LS-normal.                  & 100 & 95.65 $\pm$ 0.43 & 0.9582 $\pm$ 0.0061 & 0.9120 $\pm$ 0.0116 & 0.9379 $\pm$ 0.0052 & 0.9920 $\pm$ 0.0011 \\
                                                                             & TD-unnormal.                & 100 & 95.60 $\pm$ 0.48 & 0.9551 $\pm$ 0.0088 & 0.9133 $\pm$ 0.0104 & 0.9358 $\pm$ 0.0062 & 0.9915 $\pm$ 0.0013 \\
                                                                             & PL-unnormal.                & 100 & 95.36 $\pm$ 0.45 & 0.9602 $\pm$ 0.0098 & 0.9029 $\pm$ 0.0118 & 0.9350 $\pm$ 0.0058 & 0.9914 $\pm$ 0.0014 \\
                                                                             & LS-unnormal.                & 100 & 95.28 $\pm$ 0.39 & 0.9426 $\pm$ 0.0072 & 0.9144 $\pm$ 0.0079 & 0.9292 $\pm$ 0.0050 & 0.9898 $\pm$ 0.0011 \\ \hline
\multirow{3}{*}{\shortstack{Ensemble\\Learning}} & Stacking                    & \textbf{100} & 97.62 $\pm$ 0.25 & 0.9691 $\pm$ 0.0068 & 0.9582 $\pm$ 0.0083 & 0.9649 $\pm$ 0.0033 & 0.9962 $\pm$ 0.0007 \\
                                                                             & Hard- voting                 & 94.13 $\pm$ 0.58 & 99.04 $\pm$ 0.16 & 0.9898 $\pm$ 0.0043 & 0.9806 $\pm$ 0.0047 & 0.9842 $\pm$ 0.0026 & 0.9927 $\pm$ 0.0020 \\
                                                                             & Soft- voting                 & 92.19 $\pm$ 0.77 & \textbf{99.42 $\pm$ 0.16 } & \textbf{0.9926 $\pm$ 0.0032} & \textbf{0.9894 $\pm$ 0.0052} & \textbf{0.9885 $\pm$ 0.0039} & \textbf{0.9981 $\pm$ 0.0008} \\ \hline
\end{tabular}
\begin{flushleft}
		\textsc{Note.}\\ {
		Class. Rate is the classification rate defined in Equation \ref{equ:classrate}. The accuracy, recall, and precision are defined in Equations (\ref{equ:accuracy}), (\ref{equ:recall}), and (\ref{equ:presicion}), respectively}. The BEP and AUC are introduced in Section \ref{sec:six cnns}. The mean and standard deviations of each property are listed in the table. For each parameter in the six CNNs and ensemble-learning methods, the best results are shown in bold. TD-normal. is the best CNN among the six CNNs. Soft voting has better results except for the Class. Rate, which we discussed in Section \ref{sec:comparing cnns}. \\
	\end{flushleft}
\end{table*}


%

\end{document}